\documentstyle[aps,prb,epsf,preprint]{revtex}
\draft
\tighten
\begin{document}
\columnsep -.375in

\title{Non-Fermi Liquids in the Extended Hubbard Model}
 
\author{Qimiao Si}
 
\address{Department of Physics, Rice University, Houston,
TX 77251-1892, USA}

\maketitle 
 
\begin{abstract}
I summarize recent work on non-Fermi liquids within certain generalized
Anderson impurity model as well as in the large dimensionality ($D$)
limit of the two-band extended Hubbard model. The competition between
local charge and spin fluctuations leads either to a Fermi liquid
with renormalized quasiparticle excitations, or to non-Fermi
liquids with spin-charge separation.
These results provide new insights into the phenomenological similarities
and differences between different correlated metals.
While presenting these results, I outline a general strategy
of local approach to non-Fermi liquids in correlated electron systems.
\end{abstract}
 
\vskip 1 in

\newpage

\section{\bf Introduction}
\label{sec:intro}

Since its introduction four decades ago, Landau's Fermi Liquid theory
has been the standard model for interacting
many-fermion systems\cite{Landau}. The theory postulates that
at low energies only the quasiparticle excitations play
an essential role. The quasiparticles, essentially dressed fermions,
can be adiabatically connected to certain non-interacting 
fermions as we turn off the interaction strength. The Fermi liquid
description has been 
successful not only for the weakly interacting electrons in simple 
metals, but also for strongly correlated fermion systems. 
In this latter category are liquid ${\rm ^3He}$\cite{Wheatley,LeggettHe3}
for which the Fermi liquid theory was initially formulated, 
and the metallic states of $V_2O_3$ based 
compounds\cite{McWhan,BrinkmanRice}, the prototype material
displaying the Mott transition phenomenon\cite{mott}. 
Also included are the ``conventional'' heavy fermions\cite{Steglich,Leehf}
such as $\rm CeCu_6$ and $\rm UPt_3$, in which the mass of the
quasiparticles are enhanced by as much as hundreds from 
the band-theory predictions.

In recent years, a number of strongly correlated materials have 
emerged which show physical properties anomalous in the context of 
the canonical Fermi liquid theory. These include, in addition to
the much studied high $\rm T_c$ copper oxide 
superconductors\cite{Houston}, a class of novel heavy
fermions\cite{Maple}, $d-$ or $f-$ electron based metals
close to quantum criticality\cite{Lonzarich,Lohneysen},
quasi-one-dimensional materials\cite{Schulz,Voit}, 
as well as certain artificially fabricated metallic 
point contacts\cite{Ralph,Altshuler}. 

The theoretical question, then, is: under what conditions do electron 
correlations lead to a breakdown of Fermi liquid theory?
In the past few years, several theoretical approaches have 
been taken to address this question.  One approach builds on our 
understanding of the breakdown of Fermi liquid theory in one dimension.
For weakly interacting one dimensional fermion systems, the
perturbative renormalization group (RG) leads to the g-ology
classification of spatially homogeneous metallic states.
The possible states are Luttinger liquids or those with divergent 
CDW, SDW or superconducting correlation
functions\cite{Solyom,Haldane1D}.
In dimensions higher than one, perturbative RG analysis has 
shown that, the Fermi liquid theory does describe weakly 
interacting fermion systems with a regular density
of states\cite{Shankar,Castellani,Engelbrecht}. 
The mechanism for the breakdown of Fermi liquid theory is
necessarily non-perturbative in 
interaction strength\cite{Anderson}.

An alternative approach to non-Fermi liquids uses local
physics as a starting point. The motivations behind this
approach are multi-fold. First of all, most of the correlated 
electron systems are transition-metal, rare earth or actinides
based compounds. The dominant electron-electron interactions
in these systems are local in space. This is the result of 
quantum chemistry: the partially-filled $d-$ or $f-$ orbitals
are much more contracted than the $s-$ and $p-$orbitals
of the simple metals and covalent semiconductors, making
the intra-site Coulomb interactions by far the largest
interaction parameter. Secondly, Anderson- and Kondo- like
impurity models have been studied extensively for more than 
three decades\cite{Hewson}. In particular, the multi-channel
Kondo problem has long been recognized to display RG fixed
points of the non-Fermi liquid variety\cite{Blandin}.
Finally, the large $D$ dynamical mean field 
theory\cite{MetznerVollhardt,largeD.reviews} opens the 
door for systematic treatment of the competition between
local dynamics and spatial fluctuations.

The work summarized here covers a specific source of local physics
towards non-Fermi liquids, namely the competition between local 
charge (valence) fluctuations and spin fluctuations.
This belongs to the domain of mixed valence physics, a classic
problem in condensed matter theory. It is different from
the multi-channel physics for non-Fermi
liquids\cite{Blandin,Cox}.
In the remainder of this section, I introduce the problem, define
the models, and summarize the essential new results. More detailed
discussions of the underlying physics are given in the
subsequent sections: Sections II and III focus on the 
single impurity generalized Anderson model, and Sections IV and
V discuss the two-band extended Hubbard lattice model.

\subsection{Phenomenological considerations}

Extensive studies on the heavy fermion metals have led to a 
canonical picture for the formation of a Fermi liquid in
metals with strong local electron-electron interactions. 
Fig. 1 illustrates the point. Plotted here are the temperature
dependence of the Cu-site NMR relaxation rate\cite{Kitaoka}
(${1 /T_1}$) and that of the electrical resistivity\cite{Penney}
in $\rm CeCu_6$, one of the so-called ``vegetable'' heavy fermions.
At asymptotically low temperatures, the NMR relaxation
rate\cite{form-factor} is linear in temperature,  while
the electrical resistivity is quadratic in temperature.
Both are characteristic of quasiparticle contributions.
Simplistically speaking, the number of thermally excited spin 
excitations is proportional to temperature as a result of the 
Fermi-Dirac distribution. And each spin excitation contributes
a temperature-independent term to the flipping rate of 
nuclear spins, but a $T-$linear term to the quasiparticle 
scattering rate. The latter is again due to the Fermi statistics,
which reduces the phase space for quasiparticle-quasiparticle 
scatterings. The experimental data behave very differently
at temperatures above about $5-10$K. Here, the NMR relaxation rate
becomes essentially temperature-independent, while the temperature
dependence of the electrical resistivity is characteristic of
Kondo-scattering from local moments\cite{Hewson}. A local moment
picture serves as a better starting point to describe the
$f-$electron degrees of freedom in this temperature range.

This crossover phenomenon provided the phenomenological basis 
for the canonical theoretical picture for heavy fermion metals.
In this picture, the $f-$electrons cross over from
incoherent moments at high temperatures to being part of
the renormalized quasiparticles in a coherent Fermi liquid 
at the lowest temperatures. This is the lattice analog of the
broad crossover that is known in the solution to the single
impurity Kondo problem\cite{YuvalAnderson,Wilson,AndreiWiegmann}.
In the single impurity problem, the characteristic crossover
temperature is the Kondo energy. In the lattice case,
the crossover temperature relates to the coherence energy scale
below which the elementary excitations are
quasiparticles with heavy mass. 
The coherence energy acts as the renormalized Fermi energy for 
the low energy quasiparticle excitations.

This canonical picture appears to break down for a set 
of novel $f-$electron materials\cite{Maple}. At low temperatures,
these materials typically have an electrical resistivity linear
in temperature, accompanied by anomalous features in a host of other
physical properties. The precise mechanisms for these low temperature
non-Fermi liquid phenomenologies are at this stage unclear.
We refer the readers to the contribution of Miranda 
et al.\cite{Miranda} in this volume for a survey of theoretical
ideas. The crossover from high to low energies in these systems 
are only beginning to be addressed\cite{Aronson}. 

In the normal state of the high ${\rm T_c}$ cuprates, the spin 
dynamics appear to show a crossover qualitatively similar to 
that of the heavy fermions. 
Shown in Fig. 2 are the temperature 
dependences of the NMR relaxation rate\cite{Imai} and electrical 
resistivity\cite{Batlogg} in the optimally doped $\rm La_{2-x}Sr_xCuO4$. 
The temperature dependence of the NMR relaxation
rate behaves in a way reminiscent of that of $\rm CeCu_6$. 
It has the asymptotic low temperature $T-$linear behavior, 
crossing over to an essentially temperature-independent 
behavior at high temperatures. The crossover temperature 
scale is about $300 K$, much higher than that of the heavy 
fermions. The spin excitations can be thought of as the 
quasiparticle-quasihole continuum at low temperatures, but as
excitations derived from local moments with short range
antiferromagnetic coupling at high temperatures. This picture
is corroborated by the neutron scattering results. 
The low energy incommensurate peaks are most naturally 
accounted for in terms of a quasiparticle contribution,
while the high energy broad background
is naturally interpreted in terms of short range local moment
correlations\cite{Hayden96,SKLL}. 

However, when it comes to the temperature dependence of the
electrical resistivity, the analogy with the heavy fermions 
stops. As can be seen in Fig. 2(b), the electrical resistivity 
is linear in temperature over essentially the entire
temperature range.

By now, there exist strong evidences that the dominant contribution
to the electrical resistivity in the high $T_c$ cuprates comes
from electron-electron scattering\cite{Bonn,Ong}, as is the case
for the heavy fermions. The NMR relaxation rate is of course 
dominated by the electronic contributions. It is therefore
quite unusual that electrons in these systems yield very 
similar magnetic responses (albeit with different energy scales),
but entirely different charge transport properties.
At the microscopic level, both the heavy fermions and copper 
oxides can be described by a model with a strongly correlated 
band and a weakly correlated one. The strongly correlated band 
is formed from the $f-$ orbitals in the heavy fermions
and from the $3d_{x^2-y^2}$ orbitals in the cuprates.
The weakly correlated band is from the non$-f$ orbitals
in the heavy fermions and from the oxygen $2p$ orbitals
in the cuprates. For heavy fermions, the point of departure
for most theoretical work is the Anderson lattice model.
For cuprates, the model that serves as a sufficiently
general microscopic starting point is the three band 
extended Hubbard model\cite{Emery,VSA}. Each (planar) unit cell
contains one copper $3d$ orbital and two oxygen $2p$ orbitals.
The non-bonding combination of the $2p$ orbitals is not
expected to play an important role. When this non-bonding 
combination is ignored, the three band extended Hubbard model
reduces to a two band Anderson lattice like model.
We will call this class of models as the two band extended Hubbard
model\cite{note:ehm}.

Inspired by these considerations, the theoretical question we ask is: 
can metallic non-Fermi liquids occur in the two band extended Hubbard model?
We will address this question by incorporating general local 
interactions allowed by symmetry. Our goal is to treat interactions 
in a non-perturbative fashion, and seek to classify all 
the possible universality classes of this model.

Recent work on this subject can be naturally separated into two 
categories. Work in the first category concern exclusively 
the single impurity physics. We have generalized the standard
Anderson model by including all the on-site interactions allowed by
symmetry\cite{SiKotliar1,SiKotliar2,KotliarSi1,KotliarSi2}.
The authors of Refs. \cite{Perakis,Giamarchi,Yulu,Sire,Costi} 
have generalized the standard Anderson model by introducing 
additional species of screening fermions. Work in the second
category deal with the large $D$ limit of the lattice extended
Hubbard model\cite{SiKotliar1,SiKotliar2,KotliarSi1,SRKR,Smith}.
Here we address the physics of the lattice model based on
our understandings of the corresponding impurity problem.

Connecting the local physics of an impurity model and a lattice
model has a long tradition. For instance, the slave boson large-$N$
approach was first constructed to describe the Fermi liquid state
of the single impurity Anderson model\cite{Coleman,Read}.
The understanding of the impurity problem set the stage for the 
slave boson large-$N$ description of the coherent Fermi liquid
state of the Anderson-lattice model\cite{AuerbachLevin,MillisLee}.

\subsection{Generalized Anderson model}

The generalized Anderson model we 
introduced\cite{SiKotliar1,SiKotliar2,KotliarSi1,KotliarSi2} is

\begin{eqnarray}
H= && E^0_d n_d+ U  n_{d \uparrow} n_{d \downarrow}
+\sum_{k\sigma} E_k c_{k\sigma}^{\dagger}c_{k \sigma} +
\sum_{\sigma} t ( d^{\dagger}_{\sigma} c_{\sigma} + h.c. )\nonumber\\
&&+ {V } n_d n_c  + {J} \vec{S}_d \cdot \vec{s}_c
\label{hamiltonian.gam}
\end{eqnarray}
Here, $n_{d \sigma} = d_{\sigma}^{\dagger} d_{\sigma}$,
$n_{d } = \sum_{\sigma} n_{d \sigma}$,
$n_{c } = \sum_{\sigma} c_{\sigma}^{\dagger} c_{\sigma}$,
$\vec{S}_{d} = (1/2)\sum_{\sigma,\sigma'} d_{ \sigma}^{\dagger}
\vec{\tau}_{\sigma \sigma '} d_{\sigma'}$, with $\tau_x$, $\tau_y$, and
$\tau_z$ being the Pauli matrices, and 
$\vec{S}_{c} = (1/2)\sum_{\sigma,\sigma'} c_{ \sigma}^{\dagger}
\vec{\tau}_{\sigma \sigma '}c_{\sigma'}$.
The first four terms describe the standard Anderson model.
For the single impurity spin$-{1\over 2}$ $d-$orbital, the
energy level is  $E_d^0$, and the on-site Coulomb repulsion $U$.
For the most part, we will consider only the $U=\infty$ limit.
For the spin$-{1\over 2}$ conduction $c-$electrons, the energy
dispersion is $E_k$. 
$c_{\sigma}^{\dagger}={1\over \sqrt{N_{site}}} 
\sum_k c_{k\sigma}^{\dagger}$ is the Wannier orbital of the 
$c-$electrons at the impurity site. It hybridizes with
the $d-$electron through the hybridization matrix $t$. 

The last two terms are additional interaction terms allowed by symmetry.
$V$ describes a local density-density interaction between the 
impurity $d-$ and local $c-$ electrons. In the heavy fermion
literature, this term is called the Falicov-Kimball 
interaction\cite{FalicovKimball}. $J$ describes the
spin exchange interaction between the $d-$ and $c-$ electrons.
It describes the sum of the direct exchange interaction and the 
indirect exchange interactions mediated by those high energy
configurations not included in the model Hamiltonian.

The standard Anderson model with a featureless conduction electron
density of states has already been solved. In the particle-hole
(p-h) symmetric case, namely for $U+2E_d^0=0$, the impurity $d-$levels
are illustrated in Fig. 3(a). For sufficiently large U, the empty
impurity configuration ($|0>$) and the doubly occupied impurity
configuration  ($|2>\equiv |\uparrow \downarrow>$) can be
eliminated through a Schrieffer-Wolff 
transformation\cite{SchriefferWolff}. The result
is the Kondo problem with antiferromagnetic exchange interaction.
The latter problem is solved by a variety of methods, including 
scaling\cite{YuvalAnderson}, numerical renormalization
group (NRG)\cite{Wilson}, Bethe Ansatz\cite{AndreiWiegmann}, 
slave boson large N method\cite{Read}, and conformal field 
theory\cite{AffleckLudwig}. The conclusion is that, the low lying 
excitations can be well described by the strong coupling Fermi
liquid fixed point. The local moment is quenched by the conduction
electron spin polarization and hence disappears from the low lying
excitation spectrum.

In the p-h asymmetric case, $U+2E_d^0 >> |E_d^0|$, three
impurity configurations have to be retained at low energies. 
This is the mixed valence problem. It differs from the Kondo problem 
in that low lying  local  charge (valence) fluctuations coexist with
spin fluctuations. Historically, a variational study by Varma 
and Yafet\cite{VarmaYafet}, and RG studies of Haldane\cite{Haldane}
and Krishnamurthy {\it et al.}\cite{Krishnamurthy}, the Bethe Ansatz 
solution\cite{AndreiWiegmann} and the slave boson large-N
results\cite{Coleman} have all found that the low energy behavior
of the mixed valence problem is described by a strong coupling,
Fermi liquid fixed point. This fixed point is qualitatively similar
to that of the Kondo problem, though quantities such as the 
Wilson ratio are modified.

In Refs. \onlinecite{SiKotliar1,SiKotliar2,KotliarSi1,KotliarSi2},
we studied the generalized Anderson impurity model by 
extending Haldane's RG scheme such that the local charge
fluctuations and local spin fluctuations are treated on an 
equal footing. In the mixed valence regime, there exist three,
and only three, kinds of fixed points. In addition to the
aforementioned strong coupling Fermi liquid phase, there
are two non-Fermi liquid phases which we termed the weak 
coupling phase and the intermediate phase. The strong
coupling and weak coupling phases are the direct
analog of the strong coupling phase of the antiferromagnetic
Kondo problem and the weak coupling phase of the
ferromagnetic Kondo problem, respectively. As for the local moment case,
the weak coupling phase of the mixed valence problem requires that
the exchange coupling be ferromagnetic. Therefore,
this state is likely be of only academic value for the most part.
The possible exception is the double-exchange model for
the Perovskite Manganese Oxides\cite{manganites}.
The intermediate phase is unique to  the mixed valence regime. 
Its existence came as a surprise. 
In this new phase, spin and charge excitations are separated; 
the spin susceptibility remains to
have the Fermi liquid form as in the strong coupling phase, 
while the charge susceptibility and the single particle Green's
function have an algebraic behavior with interaction-dependent exponents.
Our RG results are substantiated by the strong coupling atomic
analysis\cite{SiKotliar2} and by the exact solutions at certain
exactly soluble points (Toulouse points)\cite{KotliarSi2}.

The single impurity model that Perakis et al.\cite{Perakis}
studied using NRG is defined as follows,

\begin{eqnarray}
H= && E^0_d n_d+ U n_{d \uparrow} n_{d \downarrow} 
+\sum_{k\sigma} E_k c_{k\sigma}^{\dagger}c_{k \sigma}
+\sum_{\sigma} t ( d^{\dagger}_{\sigma} c_{\sigma} + h.c. )\nonumber\\
&&+ {V} n_d n_c + \sum_{l=1}^N V_l n_d n_{cl}
+ \sum_{l=1}^N \sum_{k\sigma} E_k c_{l,k\sigma}^{\dagger}c_{l,k \sigma}
\label{hamiltonian.perakis}
\end{eqnarray}
where $c_{l,k\sigma}^{\dagger}$, for $l=1, ..., N$, describe 
fermionic bands that interact, but do not hybridize, with the 
impurity $d-$electron. The screening interactions $V_l$ are
introduced so that the Friedel sum rule is
satisfied\cite{Haldane.srn}. Ref. \onlinecite{Perakis} reported
NRG results in the mixed valence regime (with $U=\infty$).
The numerical results for the case of sufficiently large values
of the screening interactions were interpreted as displaying
divergent charge and spin susceptibilities 
near the mixed valence point. Such a phase is not 
expected from the Coulomb gas RG analysis. Within the
Coulomb gas RG picture, the effect of screening
fermions is to modify the initial conditions
of the RG flow\cite{SiKotliar1,Giamarchi,Hakim}.
The additional screening channels, while increasing the
orthogonality effects, do not participate in the formation
of fixed points other than those within our RG classification.
For sufficiently strong $V_l$, the Coulomb gas analysis
predicts that the mixed valence state is the intermediate
phase, with divergent charge susceptibility but {\it regular}
spin susceptibility. While further NRG studies are clearly
called for, here we note that the existing numerical
data of Ref. \onlinecite{Perakis}
might actually not be inconsistent with our Coulomb gas
RG prediction. The reason is simple. Unlike for the
charge susceptibility, the spin susceptibility continues
to increase when $E_d^0$ is decreased through the transition
regime. A true divergence is therefore hard to detect
numerically. 

\subsection{Extended Hubbard model}

The two-band extended Hubbard model is the lattice analog of the
generalized Anderson model Eq. (\ref{hamiltonian.gam}),
and is defined by the following Hamiltonian,

\begin{eqnarray}
H = && \sum_{i} {\epsilon}^o_d n_{di}
+ {U } \sum_i n_{d i \uparrow}  n_{d i \downarrow}
+\sum_{ij,\sigma} t_{ij} c^{\dagger}_{i\sigma} c_{j\sigma} 
+\sum_{i \sigma} t ( d^{\dagger}_{i\sigma} c_{i\sigma} + h.c. )
\nonumber\\
&& + \sum_{i} ( V n_{di} n_{ci} +{J} \vec{S}_{d i} \cdot \vec{s}_{c i})
\label{hamiltonian.ehm}
\end{eqnarray}
The notations are essentially the same as in 
Eq. (\ref{hamiltonian.gam}).
The only difference is that we have used ${\epsilon}^o_d$ 
to label the $d-$electron level to emphasize the difference 
of this quantity in the lattice model with that of the
impurity model $E_d^0$ (see the discussions around
Eq. (\ref{eff.dlevel}) below).

The first four terms describe the standard Anderson lattice model.
The spin-${1 \over 2}$ $d-$ electrons have an infinite
on-site Coulomb repulsion $U$, and an energy level $\epsilon_d^0$.
$t_{ij}$ describes the kinetic energy term of the
$c-$electrons. At every site, the $d-$ and $c-$ electrons hybridize
with each other through the hybridization matrix $t$.
The last two terms represent the on-site density-density and
exchange interactions between the two bands at every site. 

Unlike for the single impurity problem, most of the studies on the Anderson
lattice model in literature focus on the p-h asymmetric
case. This is because the conventional heavy fermion metallic states 
are formed only in the p-h asymmetric case (the p-h symmetric 
case has received renewed interests due to the new developments 
on the so-called Kondo insulators\cite{AeppliFisk}).
We do not know as much about the standard Anderson lattice model 
as about the standard single impurity Anderson model. Only few of
the theoretical methods that have been used in the single
impurity problem are generalizable to the lattice case.
This includes the Gutzwiller-like variational 
wavefunctions\cite{UedaRice,Brandow} and the slave boson 
large-N method\cite{AuerbachLevin,MillisLee}. All these
approaches have lead to the conclusion that the low energy regime is
described by a Fermi liquid with heavy mass quasiparticles.
The slave boson large N method has also been applied to
the copper oxide model\cite{Grilli}. In the metallic states
without long range order, the solution is again a Fermi liquid.

Our new understandings of the impurity physics, combined with the 
large $D$ approach, have led to the conclusions that metallic
non-Fermi liquid solutions are possible in the extended Hubbard
model. This conclusion is firmly established in the large $D$
limit\cite{SiKotliar1,SiKotliar2,KotliarSi1,SRKR,Smith}.
In fact, it turns out that the mixed valence condition
is much easier to realize in the lattice models than in the
single impurity model.

\section{Non-Fermi liquids in the generalized Anderson model: 
renormalization group}

We focus first on the single impurity problem.
Given that the on-site repulsion is taken to be infinity,
the generalized Anderson model can be thought of as a three
level system with a particular form of symmetry breaking.
The schematic picture is given in Fig. 4.
The three levels correspond to the three impurity configurations:
$|\alpha> = |0>, |\uparrow>, |\downarrow>$. The hybridization $t$,
density-density interaction $V$, and spin-exchange $J$ couple
these three levels to the free conduction electron bath.
Among the three levels, $|\uparrow>$ and $|\downarrow>$ are degenerate 
in the absence of external magnetic field. No symmetry, however, 
dictates the degeneracy of $|0>$ with $|\uparrow>, |\downarrow>$.
It is therefore necessary to keep track of the energy level
difference, $E_d^0$, between $|\uparrow>, |\downarrow>$ and $|0>$.
This section summarizes the RG analysis on the three-level system
near its criticality. Analysis of certain exactly-soluble points 
-- the Toulouse points -- is the topic of the next section.

\subsection{Coulomb gas representation and renormalization group}

The RG analysis is based on a Coulomb gas 
representation\cite{YuvalAnderson,Haldane} of the three-level
system. This is carried out through an expansion in
terms of the hopping amplitudes between the three configurations
of the impurity problem. This is an extension of the classic work
of Haldane\cite{Haldane} such that the local charge
fluctuations and spin fluctuations are treated on an equal footing.

In practice, it is convenient to break the exchange coupling
$ {J} \vec{S}_d \cdot \vec{s}_c$ into 
$ {J_z} S_d^z s_c^z + (J_{\perp} /2) (S_d^{+} s_c^{-} + S_d^{-} 
s_c^{+})$ where $J_{z}$ and $J_{\perp}$ represent the longitudinal and 
spin-flip components, respectively. $V$ and $J_z$ are diagonal in the 
impurity configuration basis. Their effects are to cause
different scattering
potentials for the conduction electrons when they see different impurity
configurations. When the impurity configuration is frozen in $|\alpha>$, 
the scattering potential that the conduction electron of spin $\sigma$ 
experiences is $V_{\alpha}^{\sigma}$: 

\begin{eqnarray}
V_\sigma^\sigma &&=V + {J_z/4} \nonumber\\
V^{\bar{\sigma}}_{\sigma} &&= V -{J_z/4} \nonumber\\
V_{0}^{\sigma} &&=0
\label{potentials}
\end{eqnarray}
Quite differently, the effects of $t$ and $J_{\perp}$ terms are 
to cause quantum transitions between different
impurity configurations. Specifically, the hybridization $t$ 
term causes transitions between the empty and singly occupied 
impurity configurations, and the spin-flip $J_{\perp}$ term 
the spin up and spin down impurity configurations.

To construct the Coulomb gas representation,
we expand the partition function in terms of $t$ and $J_{\perp}$,
and integrate out the conduction electron degrees of freedom. 
The resulting form for the partition function is a summation 
over histories,

\begin{eqnarray}
{Z \over Z_0}
= \sum_{\rm history} \exp (-S[{\rm history}])
\label{sumoverhis}
\end{eqnarray}
where $Z_0$ is the partition function of the free conduction electron
sea. A history corresponds to a sequence of quantum mechanical 
hopping from one impurity state to another along the imaginary time
axis. A transition from impurity state $|\alpha>$ to $|\beta>$ is
called a kink $(\alpha ,\beta)$. A history can be specified using
the notation $[\alpha_1,...,\alpha_n;\tau_1, ..., \tau_n]$ which
specifies an $(\alpha_i, \alpha_{i+1})$ kink at time $\tau_i$. 
Fig. 5 illustrates a particular history.
The statistical weight for a given history turned out to be

\begin{eqnarray}
S [\alpha_1,...,&&\alpha_n;\tau_1, ... , \tau_n]
= - \sum_i ln (y_{\alpha_i\alpha_{i+1}})
+\sum_{i}h_{\alpha_{i+1}} {(\tau_{i+1}-\tau_i) \over \xi_0}
~~~~~~~~~~~~~~~~~~~~~~~~~~~~~~\nonumber\\
&&
+ \sum_{i<j} [K(\alpha_i, \alpha_j) + K(\alpha_{i+1}, 
\alpha_{j+1})- K(\alpha_i, \alpha_{j+1}) - 
K(\alpha_{i+1}, \alpha_{j}) ]
ln {(\tau_j - \tau_i) \over \xi_0}
\label{hisaction}
\end{eqnarray}
where $\xi_0 \sim \rho_0$ is the ultraviolet inverse energy cutoff.

This action has the form of a Coulomb gas with two distinctive
species of ``Coulomb charges''. The two ``Coulomb charges''
correspond to the charge kink and spin kink,
respectively. The fugacities of the two ``Coulomb charges'' are,

\begin{eqnarray} 
y_t \equiv && y_{0,\sigma} =t\xi_0\nonumber\\
y_j \equiv && y_{\uparrow, \downarrow} = {J_{\perp} \over 2 } \xi_0
\label{fugacities}
\end{eqnarray}
As illustrated in Fig. 6, the charge fugacity $y_t$ corresponds to the 
hopping amplitude between two local states with different charge
quantum numbers. Likewise, the spin fugacity $ y_j $ describes 
the hopping amplitude between the $|\uparrow>$ and $|\downarrow>$
configurations. The fields $h_{\alpha}$ describe the energy splittings 
among the three configurations. In the absence of an external
magnetic field, $h_0= - {2 \over 3} E_d^0\xi_0$ and $h_{\sigma}
= {1 \over 3} E_d^0\xi_0$. 

The logarithmic interactions between the hopping events reflect the
reaction of the electron bath towards the 
changes of the impurity configurations. The interaction strength 
is characterized by the stiffness constants, 
$\epsilon_t= -K(0,\sigma)$ and $\epsilon_j=-K(\uparrow,\downarrow)$
which in turn are determined by the bare interaction strength
of the original Hamiltonian.
Specifically,

\begin{eqnarray}
\epsilon_{t} = &&\frac{1}{2} [( 1 - \frac{\delta^{\sigma}_{\sigma} -
\delta^{\sigma}_{0}}{\pi})^2 + (\frac{\delta^{\bar{\sigma}}_{\sigma} -
\delta^{\bar{\sigma}}_{0}}{\pi} )^2] \nonumber\\
\epsilon_{j} = &&(1 - \frac{\delta^{\sigma}_{\sigma} -
\delta^{\bar{\sigma}}_{\sigma}}{\pi})^{2}
\label{epsilontj}
\end{eqnarray}
where $\delta_{\alpha}^{\sigma}=\tan^{-1}(\pi \rho_0 V_{\alpha}^{\sigma})$
is the scattering phase shift that the conduction electron bath
of spin $\sigma$ --whose density of states is $\rho_0$ --
experiences when the impurity configuration is frozen in $|\alpha>$.

The RG equations describe how the fugacities, stiffness constants,
and the symmetry breaking field flow as we increase the 
cutoff $\xi$. We follow the formalism of Cardy\cite{Cardy}.
A detailed derivation can be found in Refs. \onlinecite{SiKotliar2}. 
Here we quote the results,

\begin{eqnarray}
&&d y_t /d ln \xi =
(1- \epsilon_t)y_t + y_t y_j\nonumber\\
&&d y_j / d ln \xi =
(1- \epsilon_j)y_j + y_t^2\nonumber\\
&&d \epsilon_t / d ln \xi =
-6\epsilon_ty_t^2
+\epsilon_j (y_t^2 - y_j^2)\nonumber\\
&&d \epsilon_j / d ln \xi
= -2\epsilon_j (y_t^2+2 y_j^2)\nonumber\\
&& d {E_d\xi} / d ln \xi = (y_t^2 - y_j^2 )
+E_d \xi (1-3 y_t^2)
\label{scaling}
\end{eqnarray}
The RG flow of the fugacities determine how the amplitude 
for making transitions between different impurity 
configurations are modified as we go towards longer
time scales. When the amplitude grows the system is 
a Fermi liquid in analogy to the formation of Fermi liquid 
in the usual Kondo problem. When this amplitude
renormalizes to zero, quantum coherence
is destroyed and Fermi liquid theory breaks down. 
This way, we cast the breakdown of  Fermi liquid theory
in the framework of the macroscopic quantum coherence (MQC) 
problem\cite{Leggett}. The transitions between Fermi liquid
and non-Fermi liquid phases are extensions of the well
known localization transitions in the MQC problem 
with one essential difference. Here we deal with a 
special {\it three-level} system instead 
of the canonical {\it two-level} system studied
in the MQC literature.  This leads to a richer phase
diagram that we describe below.

\subsection{Universality classes}

Solving the RG flow establishes the existence of three, and only three,
mixed valence fixed points. The phase diagram for the mixed valence
regime is specified in terms of the stiffness constants $\epsilon_t$
and $\epsilon_j$ and is given in Fig. 7.

The physical meaning of the strong coupling, weak coupling,
and intermediate phases is most transparent when
the three-level system is thought of as 
the hybrid of a two-level spin Kondo problem and 
a two-level charge Kondo problem. In the 
spin-Kondo problem, the ``Coulomb charge''
of the corresponding Coulomb gas 
corresponds to the spin-kink. A history of $n$ spin kinks is shown in
Fig. 8(a). The parameters of the Coulomb gas are the spin fugacity 
$y_j = J_{\perp}\xi_0$ and the spin stiffness constant
$\epsilon_j^{\prime} = 
[1 - (2 / \pi) \tan^{-1}(\pi \xi_0 J_z/4)]^2$.
The RG flow is well known\cite{YuvalAnderson} and is given in Fig. 9(a).
For antiferromagnetic $J_z$, i.e. $\epsilon_j<1$, the flow is towards
the strong coupling, Fermi liquid fixed point. While for ferromagnetic
$J_z$, i.e. $\epsilon_j>1$, the flow is towards a line of weak coupling
fixed points. In a weak coupling fixed point, there is an
asymptotically decoupled spin.

The charge Kondo effect describes the physics of the so-called
resonant-level model in the presence of a p-h symmetry.
The spinless version of the Hamiltonian (\ref{hamiltonian.gam}),
with $E_d^0 = 0$,
reduces to the resonant level model. 
It was realized some time ago\cite{Wiegmann,Schlottman}
that the resonant-level model can be asymptotically
mapped onto the anisotropic Kondo problem, with
the hybridization $t$ and the density-density interaction
$V$ playing the role of $J_{\perp}$ and $J_z$, respectively.
The ``Coulomb charge'' of the corresponding Coulomb
gas describes a charge kink.
A history of charge kinks is illustrated in Fig. 8(b). 
The charge fugacity is $y_t=t\xi_0$ and the charge stiffness constant is 
$\epsilon_t^{\prime} = (1/2)[1 - (2 / \pi) \tan^{-1}(\pi \xi_0 V/2)]^2$.
The RG flow is given in Fig. 9(b). The flow is towards a strong
coupling Fermi liquid fixed point for $\epsilon_t^{\prime} <1$,
but towards a line of weak coupling fixed points for
$\epsilon_t^{\prime} >1$. In a weak coupling fixed point, 
the charge degree of freedom is asymptotically decoupled.
Note that $\epsilon_t' >1$ corresponds to $-V > -V^{crit} = 
(2 / \pi \rho_0) tan[(\sqrt{2} - 1)\pi/2] $, i.e., a range of
finite attractive density-density interaction. It is the charge analog
of the ferromagnetic interaction.

With this background on the charge sector alone and the spin sector
alone, the meaning of the strong coupling phase of the mixed valence 
problem is transparent. Both the spin and charge Kondo problems are in
the strong coupling regime; rapid fluctuations between all three 
local configurations take place and the conduction electrons quench
both the charge and spin degrees of  freedom of the impurity.
It is expected that this phase is a Fermi liquid. A large-N analysis
of this regime indeed gives rise to this\cite{SiKotliar2}. So do the 
exact solutions of the two strong coupling Toulouse points (see next section).
Likewise, in the weak coupling phase, neither the local charge 
nor the local spin degrees of freedom is quenched. Both the spin 
and charge Kondo problems are in the weak coupling regime, and all three
atomic configurations decouple asymptotically at low energies.
The weak coupling phase requires that the spin-exchange interaction
be ferromagnetic. It is therefore very likely to be 
of only academic value with the possible exception of the
double-exchange model\cite{manganites}.

The unexpected phase is the intermediate phase. Here,
the local spin degrees of freedom is quenched, but the local
charge degrees of freedom is not. The charge Kondo problem is in
the weak coupling regime despite of the fact that the spin
Kondo problem is in the strong coupling regime.  
There are {\it two} local configurations carrying different
charges which are decoupled asymptotically. The RG analysis 
establishes this as an allowed phase. The alternative situation,
with the spin Kondo being in the weak coupling situation and
at the same time the charge Kondo being in the strong coupling
case, is not allowed: a relevant charge flipping (hybridization)
drives the spin flipping relevant.
The domain of attraction of the intermediate phase spans the
parameter range within $\epsilon_t > 1$ and $\epsilon_j < 1$.
The RG analysis cannot specify the precise boundary
between the strong coupling and intermediate phases;
the dashed line is only schematic.
In terms of the parameters of the Hamiltonian
(\ref{hamiltonian.gam}), with an on-site $V$ term,
this domain corresponds to a region with antiferromagnetic exchange 
interaction $J$ and attractive density-density interaction $V$.
Taking $V$ as an effective parameter this condition can be satisfied
in a variety of realistic models\cite{KotliarSi1}. Finite range
interactions also help realize these phases, as discussed in the
context of impurity models\cite{Perakis,Giamarchi} and in
lattice models\cite{Smith}(see Section V).
The transition between the different regimes is analogous
to the localization phase transition studied in the
context of the macroscopic quantum coherence problem\cite{Leggett} 
and more recently in the context of transport through 
constrictions in interacting quantum wires\cite{KaneFisher}.

There is one important difference in symmetry between the
asymmetric Anderson model and the p-h symmetric resonant level model.
The p-h symmetric resonant level model has a U(1) symmetry which ensures
that the impurity level stays at the Fermi energy of the
conduction electron sea. Equivalently, the singly occupied
configuration and the empty configuration is guaranteed to be
degenerate by symmetry. This degeneracy is responsible for
the charge Kondo effect. In the case of the asymmetric Anderson
model, no symmetry protects the degeneracy of the singly
occupied and the empty impurity configurations. Degeneracy 
can be achieved only through fine-tuning the bare impurity
level. The condition for this degeneracy is none other than
the condition for mixed valency. 

The phase diagram given in Fig. 7 applies only to the mixed
valence regime. When the mixed valence condition is not satisfied,
the impurity level is either too far below the Fermi energy
or too far above the Fermi energy. They correspond to the
local moment and empty orbital regimes, respectively. 
How far is too far away from the Fermi energy depends on
whether the corresponding mixed valence state falls in the
strong coupling, weak coupling, or intermediate domain. This
crossover from local moment, mixed valence, to the empty orbital
regimes are specified in the temperature versus impurity level
space in Fig. 10. 

Fig. 10(a) specifies the finite temperature crossover 
for the strong coupling case. At zero temperature, 
the mixed valence crossover extends over a scale of
$\sim \Delta^*$, the renormalized resonance width. 
The value of $\Delta^*$ depends, of course, on where we are
in the phase diagram Fig. 7. It is finite within
the strong coupling domain. As we approach the phase boundary to the
intermediate or weak coupling phases, $\Delta^*$ vanishes
in a Kosterlitz-Thouless fashion,

\begin{eqnarray}
\Delta^ * \approx ( \rho_0 )^{-1} 
\exp [-1/ \sqrt{\epsilon^{crit} - \epsilon}]
\label{delta*.kt}
\end{eqnarray}
where $ (\epsilon^{crit} - \epsilon )$ measures the distance from the
phase boundary. 

For the intermediate phase, $\Delta^* = 0$, and the mixed valence point is a 
zero temperature critical point. At finite temperatures,
there are three energy parameters: temperature ($T$), the
running symmetry breaking field
$\delta E_d ( T) \sim (E_d^0 -E_d^c)
-\Delta_0 (T\xi_0)^{(2\epsilon_t^*-1)}$,
and the running resonance width
$\Delta (T) \sim \Delta_0 (T\xi_0)^{(2\epsilon_t^*-1)}$
(where $\Delta_0 \approx \pi \rho_0 t^2$ is the bare resonance width).
The critical behavior associated with the mixed valence critical
point occurs when $|\delta E_d ( T)|< \Delta (T) <  T$.
This condition specifies the following crossover scale, 

\begin{eqnarray}
T'\sim  {1\over \xi_0} |(E_d^0 -E_d^c)/\Delta_0 |^{1 \over {2\epsilon_t^*-1}}
\label{critical.tline}
\end{eqnarray}
The correlation functions assume the form characteristic of the
intermediate phase at $T>T'$ for a given $E_d^0$, 
or equivalently, for a given temperature, when
$E_d^0$ is tuned to the range

\begin{eqnarray}
|E_d^0 -E_d^c| < \Delta_0 (T \xi_0)^{(2\epsilon_t^*-1)}
\label{critical.ed0}
\end{eqnarray}
The intermediate mixed valence regime is a manifestation 
of the quantum critical phenomenon in the context of quantum
impurity models\cite{Sachdev}.

\subsection{Further remarks}

Our most interesting finding in the mixed valence regime is the 
existence of a new phase, the intermediate phase. The Coulomb gas RG 
analysis provides the qualitative physical picture of the intermediate
phase: spin excitations are quasiparticle-like, and charge excitations
incoherent. However, the Coulomb gas RG analysis is not capable of
determining the precise forms of the correlation
functions for the intermediate
phase (neither for the strong coupling phase, for this matter). This calls
for alternative means through which correlation functions
can be calculated 
explicitly. In the next section, we present explicit results of the
correlation functions near several exactly soluble points.

The Coulomb gas representation is based on the dilute instanton expansion. 
The RG analysis, while non-perturbative in the stiffness constants,
is perturbative in terms of the fugacities. It is in principle
possible that additional fixed points, not captured by the
dilute instanton expansion, may occur. An example for the
latter arises in the related, though qualitatively different,
problem of tunneling through a point contact in a Luttinger
liquid\cite{KaneFisher,Furusaki}.
One way to probe the nature of the fixed points is to carry out
a strong coupling atomic analysis, in the same spirit that
Nozieres did for the usual Kondo problem\cite{Nozieres}.
Namely, one analyzes whether the couplings are stable
around the point where the couplings associated
with the relevant fugacities take infinite values.
We found that these points are indeed stable\cite{SiKotliar2},
giving some support that the Coulomb gas RG classification
of the universality classes are complete.
This analysis of course does not completely
rule out the existence of more fixed points.
This issue became even more urgent due to the bosonization work
of Refs. \onlinecite{Sire,Yulu}
which reported fixed points unexpected from the Coulomb gas
RG picture. It turned out that, as discussed in some detail
in the next section, there are some technical subtleties with
the application of the bosonization method to the mixed
valence problem. When these subtleties are taken care of,
the bosonization results become consistent with 
the Coulomb gas RG predictions.

\section{Non-Fermi liquids in the generalized Anderson model: 
Toulouse points}

There are three particular combinations of the interactions\cite{KotliarSi2}
where the model is exactly soluble\cite{Bethe}. These points in the interaction
parameter space are the mixed valence counterparts of the usual
Toulouse point of the Kondo problem\cite{Toulouse,YuvalAnderson},
and are naturally called the Toulouse points of the mixed valence problem.

We identify the possible Toulouse points using the bosonization
method\cite{Emery}. Given that the interaction occurs at
$\vec{r}=0$ only, we need to
keep only the $S-$wave component of the conduction electrons.
This $S-$wave component is defined on the radial axis,
$r \in [0, +\infty)$, and can be further decomposed into an outgoing and
an incoming components. In a standard fashion,
we extend to the full axis, $x \in (-\infty, +\infty)$,
by retaining only one chiral component, which we denote by
$\psi_{\sigma} (x)$. We can then introduce a boson representation
for the $\psi_{\sigma} (x)$ field. At the origin,

\begin{eqnarray}
\psi_{\sigma}^{\dagger} (x=0) = F_{\sigma}^{\dagger} {1\over \sqrt{2\pi a}}
{\rm e}^{i\Phi_{\sigma}}
\label{fermiono}
\end{eqnarray}
Here, $a$ is a cutoff scale which can be taken as a lattice spacing. 
$\Phi_{\sigma}$ is the shorthand notation for $\Phi_{\sigma}(x=0)$,

\begin{eqnarray}
\Phi_{\sigma} = \sum_{q>0} \sqrt{2\pi \over q L } ( -i
b_{q\sigma}^\dagger e^{-qa/2} + i b_{q\sigma} e^{-qa/2})
\label{bosonization}
\end{eqnarray}
where $b_{q\sigma}$ and $b_{q\sigma}^{\dagger}$ are the Tomonaga
bosons, and $L$ is the length of the dimension and is taken to be
infinite in the end. An important point is that, $\Phi_{\sigma}$
depends only on the $q\ne 0$ components of the Tomonaga bosons.
In Eq. (\ref{fermiono}), the operator $F_{\sigma}^{\dagger}$,
and its adjoint $F_{\sigma}$, are the so-called Klein factors.
They should be thought of as acting on the $q=0$ sector of the
Hilbert space for the Tomonaga bosons. More precisely, the Klein
factors can be defined as the raising and lowering operators,
respectively, in such a Hilbert space\cite{Haldane1D,Heid,Neuberg}.
These operators are unitary, and anticommute among the different
spin species. Furthermore, they commute with $b_{q\sigma}$ and
$b_{q\sigma}^{\dagger}$ for $q \ne 0$ and, hence, also with
$\Phi_{\sigma}$.

The generalized Anderson model can be rewritten in the bosonized form,

\begin{eqnarray}
H =&& H_0 + E_d^0 \sum_{\sigma}X_{\sigma \sigma} +
H_{\perp t} + H_{\perp j} +H_V\nonumber \\
H_{0} = &&\frac{{v_F}}{4\pi} \int d x [(\frac{d\Phi_{s}}{dx})^{2} +
(\frac{d\Phi_{c}}{dx})^{2}]\nonumber\\
H_{\perp t} =&& \frac{t}{\sqrt{2\pi a}} 
\sum_{\sigma} [X_{\sigma 0} F_{\sigma} e^{-i(1/\sqrt{2})\Phi_{c} }
e^{-i\sigma (1/\sqrt{2})\Phi_{s}} +  H.c.] \nonumber \\
H_{\perp j} =&& \frac{J_{\perp}}{4 \pi a} 
[X_{\uparrow\downarrow} F_{\downarrow}^{\dagger}F_{\uparrow} 
e^{-i\sqrt{2}\Phi_{s}} + H.c.] \nonumber\\
H_V = && \sum_{\alpha} X_{\alpha\alpha} 
[(\frac{{\delta}_{\alpha}^s}{\pi\rho_0}) ({1 \over 2\pi})
(\frac{d\Phi_{s}}{dx})_{x=0} 
+ ({{\delta}_{\alpha}^c \over \pi \rho_0})
({1 \over 2\pi}) (\frac{d\Phi_{c}}{dx})_{x=0} ]
\label{hamgamboson}
\end{eqnarray}
where we have used $n_d = \sum_{\sigma} X_{\sigma \sigma}$,
$d_{\sigma}^{\dagger} = X_{\sigma 0}$,
$S_d^{+} = X_{\uparrow \downarrow}$, and
$S_d^{z} = (X_{\uparrow \uparrow}
-X_{\downarrow \downarrow})/2$. 
$X_{\alpha\beta}=|\alpha><\beta|$ are the Hubbard operators.
The requirement that $\alpha,\beta$ take three, and only three,
impurity configurations, $|0>$ and $|\sigma> 
= d_{\sigma} ^{\dagger}|0>$, amounts to the following constraint,

\begin{eqnarray}
X_{\uparrow \uparrow}
+X_{\downarrow \downarrow} +X_{00} =1
\label{constraintmv}
\end{eqnarray}
In Eq. (\ref{hamgamboson}), 
$\Phi_{c,s} \equiv (\Phi_{\uparrow} \pm 
\Phi_{\downarrow})/\sqrt{2}$ are the charge and spin bosons, 
respectively. $\delta^{c}_{\alpha}
\equiv \frac{1}{\sqrt{2}} \sum_{\sigma} 
\delta^{\sigma}_{\alpha}$ and $\delta^{s}_{\alpha} \equiv
\frac{1}{\sqrt{2}} \sum_{\sigma} \sigma \delta^{\sigma}_{\alpha}$.
$v_F = 1/2\pi\rho_0$ is the Fermi velocity.

The Toulouse points are derived through applying a canonical
transformation to the Hamiltonian Eq. (\ref{hamgamboson}) and
demanding that the transformed $H_{\perp t}$ and $H_{\perp j}$
have simple forms and, simultaneously, the transformed $H_V$ vanishes.
Three such Toulouse points exist. The details are given in Ref.
\onlinecite{KotliarSi2}. In the following, we only quote 
the effective Hamiltonian, and the results for 
the single particle, spin-spin, and charge-charge correlation functions,
at each of these Toulouse points.

\subsection{Strong coupling Toulouse point I}

The first Toulouse point corresponds to $\epsilon_t=0$
and $\epsilon_j=0$. According to the phase diagram (Fig. 7)
of the Coulomb gas RG analysis, this point lies deep in
the strong coupling, Fermi liquid region.

To write the effective Hamiltonian, we need to introduce
pseudoboson operators $b_{\sigma}^{\dagger}$ and $b_0^{\dagger}$
defined as follows,

\begin{eqnarray}
X_{\sigma\sigma^{\prime}} =&&
f_{\sigma}^{\dagger}f_{\sigma^{\prime}}\nonumber\\
X_{\sigma 0} = &&f_{\sigma}^{\dagger}b_0\nonumber\\
X_{00} = && b_{0}^{\dagger}b_0\nonumber\\
b_{\sigma}^{\dagger} = &&f_{\sigma}^{\dagger}
F_{\bar{\sigma}}^{\dagger}
\label{defbs}
\end{eqnarray}
Note that the pseudoboson operator $b_{\sigma}^{\dagger}$ incorporates
a Klein operator associated with the conduction electron degrees
of freedom. In terms of these operators, the constraint 
Eq. (\ref{constraintmv}) can be rewritten as
$\sum_{\sigma} b_{\sigma}^{\dagger}b_{\sigma} + 
b_{0}^{\dagger}b_{0}  = 1$. 
The effective Hamiltonian can then be conveniently written as

\begin{eqnarray}
H_{eff}^A = &&H_0 + H_{3l} +\Delta H \nonumber\\
H_{3l}=&& E_d^0 (\sum_{\sigma} b_{\sigma}^{\dagger}b_{\sigma} -
b_{0}^{\dagger}b_{0}) + \frac{t}{\sqrt{2\pi a}}\sum_{\sigma} 
(b_{\sigma}^{\dagger} b_0+ H.c.)
- \frac{J_{\perp}}{4\pi a} (b_{\uparrow}^{\dagger} b_{\downarrow} + H.c.)
\nonumber\\
\Delta H = &&({\kappa_c \over 2\pi\rho_0}) 
(\sum_{\sigma} b_{\sigma}^{\dagger}b_{\sigma} -
b_{0}^{\dagger}b_{0}) 
({1 \over 2\pi})({d \Phi_c \over dx})_{x=0}\nonumber\\
&&+({\kappa_s \over 2\pi\rho_0}) 
(\sum_{\sigma} \sigma b_{\sigma}^{\dagger}b_{\sigma} )
({1 \over 2\pi})({d \Phi_s \over dx})_{x=0}\nonumber\\
\label{effha}
\end{eqnarray}
where $\kappa_c$ and $\kappa_s$ measure the deviation from the Toulouse point. 
This effective Hamiltonian is composed of three parts: 
$H_{3l}$ is the Hamiltonian for 
the isolated three levels, $b_{\uparrow}^{\dagger} |vac>$,
$b_{\downarrow}^{\dagger} |vac>$,
and $b_{0}^{\dagger} |vac>$, where $|vac>$ denotes the vacuum state.
The $t$ and $J_{\perp}$ are transverse fields, and $E_d^0$ provides a
longitudinal field. $H_0$ describes the 
free spin and charge bosonic fields.
Finally, $\Delta H$ is the dissipative term coupling the three levels to
the bosonic bathes. The effective Hamiltonian therefore is
a three-level generalization of the two-level ``spin''-boson 
problem\cite{Blume,Leggett}. 

All the correlation functions of this three-level ``spin''-boson problem
can be calculated explicitly. The results for the single-particle Green's function
$G_d (\tau ) = - <{\rm T}_{\tau}d_{\sigma} (\tau ) d_{\sigma}^{\dagger} (0)>$,
the density-density correlation function $\chi_{\rho}(\tau)
=<{\rm T}_{\tau}n_d(\tau) n_d (0) >$, the longitudinal and
transverse spin-spin correlation functions
$\chi_{\sigma}^{zz}(\tau) =<{\rm T}_{\tau} S^z(\tau) S^z (0) >$ and
$\chi_{\sigma}^{+-}(\tau) =<{\rm T}_{\tau} S^-(\tau) S^+ (0) >$
are given as follows,

\begin{eqnarray}
G_d(\tau) && \sim {\rho_0 \over  \tau }\nonumber\\
\chi_{\sigma}^{-+}(\tau) && \sim (\frac{\rho_0}{\tau})^2\nonumber\\
\chi_{\sigma}^{zz} (\tau) && \sim ({\kappa_s \over 2\pi\rho_0h_s})^2
(\frac{\rho_0}{\tau})^2\nonumber\\
\chi_{\rho} (\tau) && \sim ({\kappa_c \over 2\pi\rho_0h_c})^2
(\frac{\rho_0}{\tau})^{2}
\end{eqnarray}
where $h_s=J_{\perp} /4\pi a$ and $h_c = t /\sqrt{2\pi a}$.
A long time $1/\tau$ behavior for the single particle Green's
function, together with the $1/\tau^2$ behavior for the two particle
correlation functions, imply that the system is a Fermi liquid.
Therefore, this Toulouse point describes the strong coupling
phase.

Unlike for the Kondo problem, keeping track of the anticommutation
relation between fermions of different spins in the boson
representation plays an essential role. Had we not included the
Klein operator in the boson representation of the fermion operator
Eq. (\ref{fermiono}), the pseudoboson operator $b_{\sigma}$
defined in Eq. (\ref{defbs}) would not include the additional
Klein operator (it would then be a pseudofermion, as
a matter of fact). The sign of
the $J_{\perp}$ term would be reversed. It can be
seen by diagonalizing
the three level atomic problem, $H_{3l}$,
that a level-crossing would arise 
as $E_d^0$ is varied. The critical value of $E_d^0$ where levels 
cross would correspond to a non-Fermi liquid critical point.
A signature for the unphysical nature of the non-Fermi liquid
critical point associated with the level crossing can be
seen by comparing the transverse and longitudinal
spin-spin correlation functions. It can be shown that
the longitudinal spin-spin correlation function has a non-Fermi 
liquid form, but the transverse spin-spin correlation function
retains the Fermi liquid form. As a matter of fact, the level
crossing would have occurred had we started from an unphysical
model with $J_z$ antiferromagnetic but $J_{\perp}$ ferromagnetic.

Within the bosonization approach, the meaning of the atomic
configurations in the canonically transformed bases is
somewhat obscure. The physical content of these configurations 
becomes transparent once we compare them with the atomic configurations
that appear in a perturbation expansion of the original Hamiltonian
in terms of $J_{\perp}/J_z$, $J_{\perp}/V$,  $t/J_z$, $t/V$, $W/J_z$,
and $W/V$. This atomic analysis is carried out in 
Ref. \onlinecite{KotliarSi2}, from which it is physically clear that 
the ground state is always a singlet no matter what the value
of $E_d^0$ is. No level crossing is expected!

\subsection{Strong coupling Toulouse point II}

This corresponds to $\epsilon_t=1/2$ and $\epsilon_j=0$. The Coulomb gas
analysis would again predict this point to be deep
in the domain of attraction of the
strong coupling Fermi liquid phase. The effective Hamiltonian is,

\begin{eqnarray}
H_{eff}^B = && H_0 + E_{d}^o \sum_{\sigma} \tilde{f}_{\sigma}^{\dagger}
\tilde{f}_{\sigma}+
t[(\sum_{\sigma}\tilde{f}_{\sigma}^{\dagger}) \eta + H.c ]\nonumber\\
&&- \frac{J_{\perp}}{4\pi a}(\tilde{f}_{\uparrow}^{\dagger}
\tilde{f}_{\downarrow} +H.c.) 
+({\kappa_s \over 2\pi\rho_0}) (\sum_{\sigma} \sigma 
\tilde{f}_{\sigma}^{\dagger}\tilde{f}_{\sigma} )
({1 \over 2\pi})({d \Phi_s \over dx})_{x=0}
\label{effhamB}
\end{eqnarray}
Here, $\eta_k^{\dagger}$ denotes a spinless conduction
electron band; it comes from refermionizing the
charge boson. 
$\tilde{f}_{\sigma}^{\dagger} = X_{\sigma 0}F_{\sigma} F_{\eta}^{\dagger}$ 
is a pseudofermion operator.
Unlike for $H_{eff}^A$, we have kept only the $\kappa_s$ term
in $\Delta H$, as the $\kappa_c$ term is not important. All the
correlation functions can once again be explicitly determined,

\begin{eqnarray}
G_d(\tau) && \sim {\rho_0 \over \tau}\nonumber\\
\chi_{\sigma}^{-+}(\tau) && \sim (\frac{\rho_0} {\tau})^2\nonumber\\
\chi_{\sigma}^{zz} (\tau) && \sim ({ \kappa_s \over 2\pi \rho_0h_s })^2 
(\frac{\rho_0}{\tau})^2\nonumber\\
\chi_{\rho} (\tau) && \sim (\frac{\rho_0}{\tau})^2
\label{transopmv2}
\end{eqnarray}
where $\kappa_s$ is again the deviation from the Toulouse point. 
Once again, the long time behavior of the single particle and two
particle correlation functions establishes the strong coupling, 
Fermi liquid nature of this Toulouse point.

Except for the change of sign in $J_{\perp}$, the effective Hamiltonian
(\ref{effhamB}) is identical to that of Refs. \onlinecite{Sire} and 
\onlinecite{Yulu}. Once again, when the Klein operators are properly
incorporated in the bosonization representation of the fermion
fields, no level crossing occurs.

\subsection{Toulouse point for the intermediate phase}
\label{sec:intmed}

This last Toulouse point occurs at $\epsilon_t=1$ and $\epsilon_j=0$.
It is not inconsistent with the Coulomb gas results that these
values of the Coulomb gas stiffnesses lie close to such a boundary
(though it is not possible to determine the precise boundary between
the intermediate phase and the strong coupling phase from the Coulomb
gas analysis).

In order to write down the effective Hamiltonian in a convenient fashion,
we need to introduce a new basis set for the three levels, 
$|A>$, $|B>$, and $|0>$. They are defined as follows,

\begin{eqnarray}
|A> &&= {1 \over \sqrt{2}} \sum_{\sigma}(-\sigma f_{\sigma}^{\dagger}
F_{\bar{\sigma}}^{\dagger} )|vac> \nonumber\\
|B> &&= {1 \over \sqrt{2}} \sum_{\sigma} (-f_{\sigma}^{\dagger}
F_{\bar{\sigma}}^{\dagger}) |vac> \nonumber\\
|0> && = b_{0}^{\dagger} |vac>
\label{defAB0}
\end{eqnarray}
where $f_{\sigma}^{\dagger}$ and $b_0^{\dagger}$ are pseudofermion and
pseudoboson operators defined in Eq. (\ref{defbs}).
In this new basis, $n_d = \sum_{\sigma} f_{\sigma}^{\dagger} f_{\sigma}
=(X_{AA} + X_{BB}-X_{00})$, $S_d^z =(1/2)\sum_{\sigma} \sigma
f_{\sigma}^{\dagger} f_{\sigma} = (X_{AB} + X_{BA})/2$,
and $X_{\uparrow \downarrow}F_{\downarrow}^{\dagger}
F_{\uparrow} = -X_{AA} + X_{BB}$. 
The effective Hamiltonian has the following form,

\begin{eqnarray}
H_{eff}^C = &&\sum_{k \sigma} E_k c_{k\sigma}^{\dagger}
c_{k\sigma} + 2t{\sqrt{\pi a}}[X_{A0}  c_{\uparrow}
c_{\downarrow} + H.c.]\nonumber\\
&& + (E_{d}^o -\frac{J_{\perp}}{4\pi a} )X_{AA} 
+ (E_d^0 + \frac{J_{\perp}}{4\pi a} ) X_{BB} \nonumber\\
&&+\frac{\kappa_c}{2\pi \rho_0} (X_{AA} + X_{BB}-X_{00})
(c^{\dagger}_{\uparrow}c_{\uparrow} + c^{\dagger}_{\downarrow} 
c_{\downarrow}) \nonumber\\
&&+\frac{\kappa_s}{2\pi \rho_0} (X_{AB} + X_{BA})
(c^{\dagger}_{\uparrow}c_{\uparrow} - c^{\dagger}_{\downarrow} 
c_{\downarrow}) 
\label{effhamC}
\end{eqnarray}

In this effective Hamiltonian, the charge sector is described by a 
genuine \underline{charge Kondo} model. $|A>$ and $|0>$ play the
role of $|\uparrow>$ and $|\downarrow>$ of the anisotropic
spin Kondo problem and should be thought of as objects
carrying charge 2 and 0, respectively. The transformed
hybridization term is the direct analog of the spin-flip
term in the anisotropic spin Kondo problem. The residual
interaction in the charge sector, ${\kappa_c \over 2\pi \rho_0}$,
is the analog of the longitudinal exchange interaction in the 
anisotropic spin Kondo problem, with the density
playing the role of the spin in the latter.
The essential difference between the charge Kondo problem in
this mixed valence context and the spin Kondo problem lies in the
symmetry-breaking field. In the latter, the spin rotational
invariance guarantees that no explicit magnetic field term will be 
generated in the absence of an external magnetic field. In our
charge Kondo problem, the p-h symmetry is
explicitly broken, and the symmetry-breaking field 
$h^{\rm charge} = E_{d}^o - \frac{J_{\perp}}{4\pi a}$
is in general non-zero. For the impurity problem, the condition
that the renormalized $h^{\rm charge}$ vanishes can be
achieved only through fine-tuning the bare $d-$level $E_{d}^o$
to a critical value $E_d^c$.

When $h^{\rm charge}=0$ is enforced, a zero temperature
quantum phase transition takes place as $\kappa_c$ is
increased through zero. The transition is characterized
by a Kosterlitz-Thouless transition in the charge
sector; the spin sector is not critical. The phenomenology
of the intermediate phase is recovered  on the negative
$\kappa_c$ side, to which the remaining of this section
is devoted. Here, the charge sector is described 
by the weak coupling fixed points of the charge-Kondo problem, 
while the spin excitations by the strong coupling,
Fermi liquid-like fixed point of the Kondo problem.
A spin-charge separation takes place. 

Within the charge sector, the impurity configuration in the ground 
state is entirely $|0>$ for $h^{\rm charge}<0$, and $|A>$ for 
$h^{\rm charge}>0$. This is the result of infinite charge 
susceptibility in the corresponding ferromagnetic charge Kondo
problem. Exactly at $h^{\rm charge}=0$, namely, when $E_d^0$
is tuned to the critical value $E_d^c= \frac{J_{\perp}}{4\pi a}$,
the impurity degrees of freedom in the ground state involve
an equal, incoherent, mixture of $|0>$ and $|A>$. Schematically,
the ground state wavefunction can be written as
$\phi = |A> \phi_{A} + |0> \phi_{0}$ 
where $\phi_{A}$ and $\phi_{0}$ are the wave functions of the
conduction electrons such that $\phi$ is the solution to
a ferromagnetic Kondo model with zero magnetic field. 
With $h^{\rm charge}=0$, the intermediate mixed valence
dynamics applies at all temperatures. When $E_d^0$ is moved 
away from the critical value,  a finite cross-over
temperature $T_{co} \sim |E_d^0-E_d^c|$ emerges. The
intermediate mixed valence dynamics continue to apply
at $T > T_{co}$. At low temperatures ($T < T_{co}$),
however, the charge fluctuations become gapped out.

The single-particle, density-density, longitudinal and transverse
spin-spin correlation functions are given as follows,

\begin{eqnarray}
G_d(\tau) && \sim (\frac{\rho_0}{\tau})^{[{1 \over 2}+{1\over 2}(1-\sqrt{2}
{\kappa_c \over \pi})^2]}\nonumber\\
\chi_{\rho}^c (\tau)
&&\sim \frac{(\rho_0t)^2}{(-4\kappa_c)}
({\rho_0 \over \tau})^{(-4\kappa_c)}\nonumber\\
\chi_{\sigma}^{-+} (\tau) && \sim (\frac{\rho_0}{\tau})^{2}\nonumber\\
\chi_{\sigma}^{zz} (\tau) &&\sim ({{\kappa_s} 
\over 2 \pi \rho_0h_s})^2 ({\rho_0 \over \tau})^2
\label{cf.intmed}
\end{eqnarray}
where $\chi_{\rho}^c$ labels the connected part. 
The exponent for the single particle Green's function is particularly
noteworthy. The ${1 \over 2}$ part is the contribution of the spin 
degrees of freedom. It is independent
of interactions, and is the same value as we would get for a non-interacting problem! 
The remaining part, ${1\over 2}(1-\sqrt{2}{\kappa_c \over \pi})^2$ is due to the charge
degrees of freedom and is interaction dependent. This is consistent with the physical
picture that in the intermediate phase the low lying spin excitations
are quasiparticle-like while charge excitations have the non-Fermi liquid form.

Other correlation functions in the charge sector also have an
algebraic behavior with interaction-dependent exponents,
and a pairing susceptibility,

\begin{eqnarray}
<{\rm T}_{\tau}(\sum_{\sigma}c_{\sigma}d_{\bar{\sigma}})(\tau)
(\sum_{\sigma}d_{\bar{\sigma}}^{\dagger}c_{\sigma}^{\dagger})(0)>
\sim (\frac{\rho_0}{\tau})^{({\kappa_c \over \sqrt{2}\pi})^2}
\label{pairingintmed}
\end{eqnarray}
is enhanced compared to the Fermi-liquid case.
This makes it plausible that the ground state in the corresponding 
lattice model is superconducting. In that case, the intermediate
mixed valence dynamics would describe the physics in the normal state,
i.e., at temperatures between the transition temperature and some
upper cutoff energy scale.

To summarize, the explicit results for the correlation functions
in this Toulouse point highlight all the features expected of an
intermediate phase: spin-charge separation; a quasiparticle
residue vanishing in a power-law
fashion; Fermi liquid like spin correlation functions; and 
self-similar local charge correlation functions with
interaction-dependent
exponents. We note in passing that these characteristics bear
strong similarity to
those of the Luttinger liquid in one dimensional interacting fermion
systems\cite{Solyom,Haldane1D,Schulz,Voit}.

\section{The extended Hubbard model as a lattice of Anderson
impurities: large $D$ limit}

We now turn to the lattice model, Eq. (\ref{hamiltonian.ehm}),
which we suggestively rewrite as

\begin{eqnarray}
H =&& \sum_i h_i + \sum_{<ij>}h_{ij} \nonumber\\
h_i = && {\epsilon}^o_d n_{di} 
+ {U } n_{d i \uparrow}  n_{d i \downarrow} 
+t ( \sum_{\sigma} d^{\dagger}_{i\sigma} c_{i\sigma} + h.c. )\nonumber\\
&& + V n_{di} n_{ci} +{J} \vec{S}_{d i} \cdot \vec{s}_{c i}\nonumber\\
h_{ij} = && t_{ij} \sum_{\sigma} c_{i\sigma}^{\dagger}c_{j\sigma}
\label{hamiltonian.ehm2}
\end{eqnarray}
This is pictorially illustrated in Fig. 11(a), in which each black dot
represents an $h_i$ term, and across each bond on the lattice there is
an $h_{ij}$ term.

\subsection{Mapping to a self-consistent impurity problem in the large $D$ limit}

The limit of infinite dimensions\cite{MetznerVollhardt,largeD.reviews}
is defined by scaling the hopping term, $t_{ij}$, 
in terms of the dimensionality ($D$) such that the limit 
is well-defined. For the nearest neighbor hopping term

\begin{eqnarray}
t_{<ij>} = {t_0 \over \sqrt{2D}}
\label{hopping.scaling}
\end{eqnarray}
The $D\rightarrow \infty$ limit is taken with $t_0$ kept fixed.

We recall that, the large $D$ limit of a classical non-frustrated
lattice spin system is taken by scaling the nearest neighbor coupling
to be of order $1/D$. For any given site, the homogeneous contributions
from neighboring sites add up to an effective field, of order unity,
that acts on the spin of the selected site. All other contributions 
are of finite orders in $1/D$ and vanishes in the large $D$ limit.
This is the content of the Weiss molecular field theory for classical
magnets. In the quantum systems, the single particle hopping contributes
to the kinetic energy of the fermion system which, at zero temperature,
is the zero point energy associated with quantum fluctuations.
The $1/\sqrt{D}$ scaling in Eq. (\ref{hopping.scaling}), as opposed to the
$1/D$ scaling, is necessary to capture these quantum fluctuations. With
this scaling, the average kinetic energy per unit cell is of order unity
in the large $D$ limit.

In finite dimensions, when on-site interactions are present, a single 
partile hopping term will generate effective intersite interactions
involving two or more particles. With the single particle hopping
term being scaled as in Eq. (\ref{hopping.scaling}), the generated 
interactions are of order $1/D$ or higher.

For a selected site, say site $0$, the effect of the rest of
the sites is to generate a retarded Wiess mean field that couples
to the single particle degrees of freedom at site zero.
The modifications to the on-site dynamics involving two or more
particles are higher order in $1/D$ and do not survive the large $D$ limit.
The result is that, all local correlation functions of the lattice model
can be entirely determined by the following effective on-site action:

\begin{eqnarray}
S^{eff}_{imp} =  S_0 -
\int_0^{\beta}&& d\tau \int_0^{\beta} d\tau' 
\sum_{\sigma} c_{0\sigma}^{\dagger}(\tau) g_0^{-1} 
(\tau-\tau') c_{0\sigma} (\tau') 
\label{action.imp}
\end{eqnarray}
$S_0$ is the action associated with $h_0$. Since $h_0$ contains
all the local interactions, the procedure treats the local interactions 
in a dynamical fashion. $g_0^{-1}(\tau-\tau') $, or equivalently,
its Fourier transform, $g_0^{-1} (i\omega_n)$, where
$i\omega_n$ is the fermionic 
Matsubara frequency, is retarded. This is the
result of integrating out the rest of the degrees of freedom 
other than site 0, at the one particle level. 
Pictorially, $g_0^{-1}$ describes the effect of all the Feynman
trajectories in which one electron leaves site zero, explores
the lattice, and returns to the origin. Translational invariance
demands that the local correlation functions of the lattice model
are site-independent, and are the same as the correlation functions
of the impurity model. This leads to the following
self-consistency equation,

\begin{eqnarray}
g_0^{-1} (i\omega_n) = -\sum_{ij}t_{i0}t_{0j}[G_{ij}(i\omega_n) - 
G_{i0}(i\omega_n)G_{0j}(i\omega_n)/G_{00}(i\omega_n)]
\label{self-consist.sp}
\end{eqnarray}
where $G_{lm} (\tau) \equiv -<T_{\tau} c_{l\sigma}(\tau)
c_{m\sigma}^{\dagger}(0)>_H$
is the lattice Green's function. 
Eqs. (\ref{action.imp},\ref{self-consist.sp}) define the dynamical mean
field formalism that is exact in the large $D$ limit\cite{largeD.reviews}.

It is physically more transparent to rewrite $S^{eff}$ in the 
Hamiltonian form. We achieve this by introducing a non-interacting 
electron bath whose dispersion and coupling to the $c_{0\sigma}$ has
to be determined self-consistently.
Introducing $\eta_{k\sigma}^{\dagger}$  and
$\eta_{k\sigma}$ as the creation and annihilation operators
for this fermion bath, where $k$ is a dummy momentum variable,
we can rewrite the effective impurity problem in terms of the
following effective impurity Hamiltonian,

\begin{eqnarray}
H^{eff}_{imp} = h_0 + \sum_{k\sigma} t_k (\eta_{k\sigma}^{\dagger} 
c_{0\sigma} + H.c.)
+\sum_{k\sigma} \epsilon_k \eta_{k\sigma}^{\dagger}\eta_{k\sigma}
\label{hamiltonian.imp}
\end{eqnarray} 
The self-consistency equation (\ref{self-consist.sp}) is equivalent to

\begin{eqnarray}
\sum_k t_k^2/(i\omega_n - \epsilon_k) = 
-\sum_{ij}t_{i0}t_{0j} [ G_{ij}(i\omega_n) - 
G_{i0}(i\omega_n)G_{0j}(i\omega_n)/G_{00}(i\omega_n)]
\label{self-consist.ham.sp}
\end{eqnarray}
Pictorially, we have reduced the task of solving the full lattice problem
of Fig. 11(a) into solving a fully interacting quantum impurity embedded in
a self-consistent fermionic sea, as is illustrated in Fig. 11(b).

\subsection{Solution to the effective impurity problem}

The crucial question is the nature of the self-consistent bath.
What we found is that, as long as the solution is metallic, 
the density of
states of the fermionic bath at the Fermi energy is {\it finite}. This is
a self-consistent statement. The reasoning goes as follows. Assuming
that the bath density of states is finite at Fermi energy, we can
proceed to solve the impurity problem in an asymptotically exact fashion
by applying bosonization technique.
Among the quantities that can be calculated asymptotically exactly is
the local $c_0$ Green's function. That this Green's function has a regular 
spectral function is seen, in the Coulomb gas representation, by noting
that the local $c_0$ does not creat a kink. A regular $c_0$ Green's
function implies a regular self-energy for the $c-$electrons, which,
through the self-consistency equation (\ref{self-consist.ham.sp}),
in turn implies that the density of states of the 
self-consistent fermionic bath is regular! The self-consistency is hence
established. This argument applies to any lattice. In the special
case of Lorentzian density of states, this statement is more than
asymptotically exact; it is exact. In the case of a Bethe
lattice with infinite coordination, we have numerically solved the
self-consistency equations\cite{SRKR}. We indeed found that
the density of states of the bath fermions at the Fermi energy 
is finite as long as the solution is metallic.
In Fig. 12, we plot the imaginary part of the $d-$electron and $c-$
electron Green's function as a function of the Matsubara frequency.
The zero frequency limit of this Green's function is
identical to the density of states at the Fermi energy.
It is clear that, even though the $d-$electron Green's function
is divergent, as expected in the non-Fermi liquid form discussed
in the previous sections, the $c-$electron density of states is regular. 

Armed with this understanding of the fermionic bath, the only
essential difference of this self-consistent Anderson model with 
the single impurity generalized Anderson model is that, this time 
the effective $d-$level is 

\begin{eqnarray}
E_d^{imp} = \epsilon_d^0-\mu 
\label{eff.dlevel}
\end{eqnarray}
instead of $\epsilon_d^0$. The effective impurity level here
is measured with respect to the Fermi energy of the lattice
model which is, of course, different for different amount
of electrons.

The fact that the density of states of the self-consistent fermionic
bath at the Fermi energy is finite implies that, we can carry out
an asymptotically exact analysis on the self-consistent
generalized Anderson impurity model exactly the way we did
for the single impurity generalized Anderson model. 
The classification of the possible phases of the single
impurity Anderson model applies to the self-consistent
Anderson impurity model, provided that we express the
stiffness parameters of the phase diagram,  Fig. 7,
in terms of the self-consistent parameters. In particular,
there is a  metallic non-Fermi  liquid state of the 
extended Hubbard model that corresponds to the intermediate
phase of the impurity model.
As in the impurity model, we have a spin excitation
spectrum that is spin-${1 \over 2}$ quasiparticle-like,
and a charge excitation spectrum that is incoherent.
What we have is a local route towards spin-charge separation
in the extended Hubbard model. 
 
The fact that the impurity level of the self-consistent
impurity problem is measured with respect to the chemical
potential implies that, for a model with fixed $\epsilon_d^0$,
the temperature-chemical potential crossover
is given in Fig. 13(a).

\subsection{Pinning of chemical potential}

A remarkable phenomenon takes place. This is the pinning of 
chemical potential\cite{skg}. The critical chemical
potential, at which the mixed-valence state persists to
zero temperature, corresponds to a range of electron
densities. It can be seen as follows. The local correlation
functions of the extended Hubbard model in infinite 
dimensions are given by the impurity problem. In particular, 
the occupation numbers of the lattice, for a given chemical
potential, can be obtained from the local Green's functions 
of the corresponding impurity model. It follows 
from our analysis of the impurity model that, at zero temperature,
$n_d$ (and also $n=n_d+n_c$) are discontinuous functions
of the chemical potential: 
$n_d \sim n_d^{+} \approx 1 - O(t^2)$ for $\mu > \mu_c$, 
while $n_d \sim n_d^{-} \approx O(t^2)$ for
$\mu < \mu_c$. At finite temperatures, 
$n_d$ is increased from $n_d^{-}$ to $n_d^{+}$
as $\mu$ is increased from $\mu_c - \Delta \mu$ to $\mu_c 
+ \Delta \mu$ where 

\begin{eqnarray}
\Delta \mu \sim \Delta_0 ( T \rho_0 )^{(2\epsilon_t^*-1)}
\label{delta.mu}
\end{eqnarray}
As long as $n_d$ is over the range $(n_d^{-}, n_d^{+})$, the condition
Eq. (\ref{critical.ed0}) is satisfied. The correlation functions will
be controlled by the intermediate phase at criticality.

It is remarkable that the fact that our impurity model is associated
to a lattice problem forces the effective impurity model to be at
criticality, with a larger symmetry than we would have naively
expected. Physically, for an incoherent state to be metallic,
it is necessary to allow charge transfer between the local degrees
of freedom and the bath. This can only happen if the local charge
degrees of freedom is in equilibrium with the conduction electron bath.
This requires the heavy level to be at the chemical potential.

\subsection{Further remarks}

The RG analysis that leads to the classification of the possible phases
of the effective impurity problem discussed in Section II is based on
a small hybridization expansion. In the context of the low energy 
effective Hamiltonians for the high $T_c$ system, one of the important
questions is whether the extended Hubbard model can be further reduced
to an effective one band Hubbard model\cite{ZhangRice,EmeryReiter}. 
One argument used in this context is that, when the hybridization is
sufficiently large, it is more appropriate to first diagonalize the
problem within a unit cell, leading to the Zhang-Rice singlet\cite{ZhangRice}.
The effective Hamiltonian for the low energy local orbitals is the
so called $t-J$ Hamiltonian. In light of this construction, a natural
question to ask is whether the physics of the extended Hubbard model
at large hybridization is different from that at small hybridization.
The large $D$ limit provides a unique opportunity to address this
question. This problem has only been numerically studied in the spinless
version of the extended Hubbard model\cite{SRKR}. The numerical
solution indicates that the qualitative phase diagram is similar
for the large and small hybridization limits. However, the
precise values of the exponents of the correlation functions could
be modified as the hybridization is increased. Further work
along this direction needs to be carried out.

We have established the existence of metallic non-Fermi liquid
states in the extended Hubbard model Eq. (\ref{hamiltonian.ehm})
in the large $D$ limit. What happens in finite dimensions?
This is a question which is only beginning to be addressed.
Some progresses are reported in the next section.

We close this section on a methodological note. Various numerical
methods, such as quantum Monte
Carlo\cite{Jarrell,Rozenberg,GeorgesKrauth}
and exact diagonalization\cite{Caffarel,SRKR},
can be used to solve the self-consistent dynamical mean field
equations associated with the $D=\infty$ limit.
Whatever the means, the solution to these equations should always
describe the solution to an impurity coupled to a self-consistent
fermionic bath. And it is always instructive to ask a) what
is the nature of the density of states of the self-consistent
fermionic bath near its Fermi energy
(is it regular, gapped, vanishing with a
power law, or singular with a power law, to name a few); and
b) what are the low lying levels associated with the impurity.
Armed with these information, it is usually possible to use
RG or other analytical means to determine the qualitative
behavior of the solution.

\section{Competition between the local and short range
fluctuations: alternative large-$D$ limit}

One major advantage of the large $D$ approach is being able 
to treat local correlations in a fully dynamical fashion.
This feature is responsible for our uncovering the metallic
non-Fermi liquids in the extended Hubbard model
that other methods have failed. One major disadvantage of the large$-D$
approach is that, spatial fluctuations beyond one particle level are
all frozen: intersite interactions reduce to Hartree contributions only.
For physical systems in finite dimensions, intersite RKKY or
Superexchange type interactions are expected to compete with local
correlations. For instance, an unstable non-Fermi liquid fixed point
arises due to the competition between the inter-impurity RKKY interaction
and the local Kondo couplings in the two-impurity
Kondo problem\cite{Jones}.
In the Kondo lattice models, such a competition led to the competition
between long range magnetic ordering and Kondo singlet formation.
In the absence of long range ordering, the dynamical role of
the intersite interactions on the local Kondo-like physics
has largely been left unexplored. From the large$-D$
point of view, one way to recover the spatial fluctuations
is the perturbative $1/D$  expansion. 
Truncating the perturbation series to order $(1/D)^n$ requires solving
at once clusters containing one, two, ...,$n+1$ sites embedded in their
respective self-consistent media\cite{Schiller}. The practicality
of this procedure is unclear at this stage. An alternative route is a
loop expansion with the requirement that the $D=\infty$ results be
recovered at the saddle-point level, as has been constructed
in models with certains forms of quenched disorder\cite{Vlad}. For clean
systems, it turns out to be difficult to formulate such a loop expansion.

We have recently introduced an alternative large $D$ limit to study
the interplay between local correlations and short range 
spatial fluctuations in the two band extended Hubbard
model\cite{Smith,Kajueter}. In this procedure, an explicit
intersite density-density interaction  term is introduced,
and is scaled in terms of the dimensionality such that its
{\it fluctuation part} survives the large D limit. This procedure
leads to an impurity embedded in a self-consistent fermionic bath
{\it and a self-consistent bosonic bath}. Detailed 
analysis\cite{Smith} has so far been carried out only for the
spinless version of the extended Hubbard model, 
given by the following Hamiltonian,

\begin{eqnarray}
H= &&\sum_{i} [E^0_d n_{di}+ t ( d^{\dagger}_i c_{i}+ h.c. ) + V n_{di}
n_{ci}]\nonumber\\
&&+\sum_{<ij>}[t_{ij} c_{i}^{\dagger}c_j + v_{ij} :n_{di}: :n_{dj}:]
\label{hamiltonian.tv}
\end{eqnarray}

The standard large $D$ limit is taken with $t_{ij}$ of the form Eq.
(\ref{hopping.scaling}) and $v_{ij}$ of order $1/D$.
 With that scaling, only the static component of
$v_{ij}$ gives a non-vanishing contribution  in the large $D$ limit. Hence,
intersite interactions give only a Hartree contribution.
The alternative large $D$ limit is taken with Eq. (\ref{hopping.scaling}) and 

\begin{eqnarray}
v_{<ij>} = v_0 / \sqrt{D}
\label{vijscaling.new}
\end{eqnarray}
In order to have a well-defined  large $D$  limit, it  is necessary that 
the zero frequency mode, i.e. the Hartree term, be treated  separately. 
In the absence of symmetry breaking, the effect of the Hartree term 
is a change to the chemical potential.  This is handled through 
normal ordering,  $:n: \equiv
n-<n>$.

The procedure outlined  in the previous section leads to the
following effective impurity  action,

\begin{eqnarray}
S^{eff}_{imp} = S_0 - \sum_{i\omega_n} c_{0}^{\dagger}(i\omega_n)
 g_0^{-1} (i\omega_n)  c_0 (i\omega_n)
-\sum_{i\nu_n \ne 0} n_{d0} (i\nu_n) \chi_0^{-1} (i\nu_n) n_{d0}(i\nu_n)
\label{seff.new}
\end{eqnarray}
In addition to the self-consistency equation (\ref{self-consist.sp}),
an additional self-consistency equation  is required, this one for
the density propagator,

\begin{eqnarray}
\chi_0^{-1} (i\nu_n) = \sum_{ij} v_{i0} v_{0j} [\chi_{ij} (i\nu_n)-
 \chi_{i0}(i\nu_n)\chi_{0j}(i\nu_n)/\chi_{00}(i\nu_n)]
\label{self-consist.new}
\end{eqnarray}
where 
$\chi_{lm} (i \nu_n)$ is the Fourier transform of the
lattice density-density correlation function,
$\chi_{lm} (\tau) \equiv <T_{\tau} :n_{l}:(\tau)
:n_{m}:(0)>_H$. $i\nu_n$ is the bosonic Matsubara frequency.

The effective action can once again be written in terms of a
single impurity Hamiltonian.
In our spinless case, this is a self-consistent
resonant-level  model  with an additional screening bosonic bath,

\begin{eqnarray}
H^{eff}_{imp} = &&(\epsilon_d^0 - \mu) n_{d0} + 
t ( d^{\dagger}_0 c_{0}+ h.c. ) + V n_{d0}
n_{c0}\nonumber\\
&&+\sum_{k} t_{k} (\eta_k^{\dagger} c_0 +h.c.) 
+\sum_k \epsilon_k \eta_k^{\dagger}\eta_k\nonumber\\
&&+\sum_{q}F_{q} (\rho_q + \rho_{-q}^{\dagger})
:n_{d0}: +\sum_q W_q \rho_q^{\dagger}\rho_q
\label{hamiltonian.imp.tv}
\end{eqnarray}
Here, $\eta_k^{\dagger}$ creats, like in the previous section, a fermionic
bath with a dummy momentum variable $k$. The dispersion, $\epsilon_k$,
 and the hybridization coupling parameter, $t_k$, are
to be determined self-consistently. Likewise, 
$\rho_q^{\dagger}$ creats a bosonic bath with a dummy momentum 
variable $q$. The corresponding self-consistent dispersion and
coupling parameters are $W_q$ and $F_{q}$.

Detailed  analysis shows that, in this case, the fermionic bath density of 
states remains regular. The spectral function of the bosonic bath,
however, can be highly non-ohmic. This is fortunate, for an impurity 
model with a fermionic bath having an arbitrary form of density of states 
near the Fermi energy is quite difficult to
handle\cite{Fradkin,Ingersent,Si}. On the other hand, an impurity
model coupled to a bosonic bath with non-ohmic spectral function
can still be analyzed asymptotically exactly, within
a modified kink-gas picture. Details  are given in Ref. \onlinecite{Smith}.
The most interesting regime is again the mixed valence regime, for
which the renormalized effective impurity level is zero. The self-consistent
solution is shown in Fig. 14. The phase diagram is specified in terms
of three parameters, $g_t=\rho_0 t$, $g_V =[1-(2/\pi)tan^{-1}
(\pi \rho_0 V/2 )]$, and $g_v=\rho_0v_0$. They are essentially
the dimensionless hybridization, on-site density-density interactions,
and intersite density-density interactions. 

At $g_v=0$, the problem reduces to the usual large $D$ case discussed in the
previous section\cite{SRKR}. For our spinless problem, a 
Kosterlitz-Thouless transition takes place describing the charge-Kondo 
effect\cite{Wiegmann}. When $g_V < g_V^{crit}$, i.e., $-V < V^{crit}_0$,
the solution is a Fermi liquid. For $g_V > g_V^{crit}$, i.e., 
$-V > V^{crit}_0$, the solution is a line of non-Fermi liquids 
with the connected local density susceptibility,

\begin{eqnarray}
\chi (\tau) \approx ( {\rho_c / \tau } )^{\alpha}
\label{susc.nfl}
\end{eqnarray}
The exponent $\alpha$ is interaction-dependent, increasing
from 0 to 2 as one moves away from the critical
point\cite{Bhatt,Imbrie}.

The intersite interaction $v_0$ modifies the phase diagram in several
ways. Consider first $g_V > g_V^{crit}$. The line of fixed points of
the $v_0=0$ problem becomes unstable. Remarkably, the correlation functions
in the new fixed points can be determined. In fact, they have the same form
as given in Eq. (\ref{susc.nfl}).

For $g_V<g_V^{crit}$, we are able to establish
the existence of a phase transition as $v_0$ is increased. 
For sufficiently strong $v_0$, the solution must be a non-Fermi
liquid. Physically, the intersite density-density
interactions provide charge-screening,
which contribute to the orthogonality
effect. In the mixed valence regime,
this orthogonality helps realize the weak coupling fixed point
with incoherent charge excitations.
For sufficiently small $v_0$, on the other hand,
the Fermi liquid solution is stable. 

As a result, non-Fermi liquids with self-similar 
correlation functions occur even for repulsive values of
the on-site density-density interaction. 
It is therefore not necessary to require attractive on-site interactions
to realize the incoherent charge state.

A finite intersite interaction $v_0$ also changes the nature of the
phase transition. We have shown that\cite{Smith} the phase transition
is not of the Kosterlitz-Thouless type. We have been unable to establish
the precise nature of the phase transition.

The self-consistent equations and the kink gas analysis can also be 
carried out in the spinful Hubbard model. The form of scaling
equations implies that the results derived here for the spinless model
carries over to the charge sector of the spin-charge separated 
intermediate phase\cite{Smith2}.

The formalism outlined in this section can be generalized to various
different contexts. The effects of intersite spin exchange interactions
are the obvious next problem to study. Three or more particle intersite
interactions can also be treated along this line.

\section{Conclusions, new insights, and open questions}
 
The focus of this review article has been on the competition between
local charge and local spin fluctuations, both in the single
impurity Anderson model and in the lattice extended Hubbard model.
We have found that such a competition leads to metallic non-Fermi liquids
in certain interaction parameter range.
In particular, we have identified a novel non-Fermi liquid 
mixed valence state, called intermediate phase.
This phase displays the phenomenon of spin-charge separation.
The low energy spin excitations are much like that of the strong 
coupling Fermi liquid phase as one would derive from, for instance,
the slave boson condensed phase within the slave boson large-N approach.
The charge excitations are distinctively of a non-Fermi liquid form.

Are these non-Fermi liquid states relevant to real materials?
For single impurity problems, these non-Fermi
liquid phases can be realized only when the impurity level is
tuned to be close to the Fermi energy of the conduction electrons.
We can envisage two contexts in which this kind of fine tuning
of the impurity level is physically feasible. The first is in the
context of dilute impurities in metals. The Fermi level of
the conduction electron sea can be varied by substituting some of the
elements in the compound with ones of different valency.
Called Fermi level tuning, this mechanism has already been
invoked to explain the trend of the Kondo energy in certain
Uranium based heavy fermions\cite{Allen,Maple.flt}. Valence fluctuations
in Uranium based compounds are in general much stronger than
in Cerium based compounds. It is conceivable that some of
them have interaction parameters that fall in the domain of
the intermediate phase. Systematic studies of the Fermi level
tuning phenomenon, therefore, can play a significant role in the
current search and study of non-Fermi liquids in
$f-$electron based materials.  The second context where
impurity level can be tuned is in  mesoscopic systems.
This time, the tuning is achieved through biasing the confined area
with respect to the leads\cite{Kastner}.

For lattice problems, the level tuning requirement is much less
stringent. This is the result of the phenomenon of the pinning
of chemical potential, discussed extensively in Sec. IV. There
is a range of electron density over which the {\it effective}
impurity level lies close to the Fermi level of the self-consistent
fermion bath. Our findings of the existence of both the strong
coupling and intermediate phases in the mixed valence regime
provide new insights into the similarities and differences
between the heavy fermions and high $T_c$ cuprates, mentioned
at the beginning of this manuscript. The spin dynamics of
the intermediate phase and the strong coupling phase are similar,
both displaying a crossover from the high temperature local moment regime
into the low temperature coherent regime.
The charge dynamics, on the other hand, are qualitatively different
in these two phases.
In the strong coupling Fermi liquid phase, the charge dynamics
track with the spin dynamics. In the intermediate phase,
the charge dynamics have non-Fermi liquid behavior
characterized by the correlation functions discussed in
Sections II and III. 

The phenomenology of the conventional heavy fermions, such as
${\rm CeCu_6}$ and ${\rm UPt_3}$, are well described by
the strong coupling Fermi liquid phase. Can the normal state
of the high $T_c$ cuprates be described by the intermediate phase?
The contrasting behaviors seen in the temperature dependences
of the NMR relaxation rate and the electrical resistivity 
in the cuprates (Fig. 2), when viewed in the context of
those of the heavy fermions (Fig. 1),
are consistent with the qualitative differences between the
spin and charge dynamics expected in the intermediate phase.
However, these two quantities measure very different correlation
functions. The NMR relaxation rate measures mainly the
momentum ($\vec q$)
averaged electron-spin response, while the electrical resistivity the
$\vec q = 0$ electrical current-current correlation function.
Therefore, the precise implications of the contrasting temperature
dependences of these two quantities in the cuprates are hard to specify.
One clear-cut probe of the relationship between the spin and
charge excitations would be to compare the temperature
dependences of the electron-spin resistivity and
the electrical resistivity\cite{Spindiff}. This requires an
experimental measurement of the electron-spin diffusion constant.

Our results also raise a number of theoretical questions. Unlike for 
the multi-channel Kondo problem, the lattice model we have studied, 
Eq. (\ref{hamiltonian.ehm}), has a well defined limit of vanishing
interactions, $U \rightarrow 0$, $V \rightarrow 0$ and $J \rightarrow 0$.
In this limit of vanishing interactions, and for dimensions higher
than one, the perturbative RG analysis\cite{Shankar} would identify
no instability towards a metallic non-Fermi liquid state.
By focusing on the strong coupling limit, $U=\infty$, and taking
the limit of infinite dimensions, $D = \infty$,
we have identified non-Fermi liquid solutions.
Obvious questions arise: a) what happens as the on-site
Hubbard interaction $U$ gradually decreases from infinity,
all the way to  $U=0$? If a phase transition takes place,
is it also of the  Kosterlitz-Thouless form?
b) What happens when the dimensionality decreases from infinity
to physical dimensions? The approach outlined in Section V provides
a starting point to address one aspect of the finite
dimensionality effects,
namely the competition between the on-site and inter-site correlations.
The results summarized there imply that the non-Fermi liquid
phases survive the short range spatial fluctuations. The critical
behavior of the quantum phase transition from the Fermi liquid
to non-Fermi liquid states, on the other hand, are strongly
modified by the spatial correlations. As for a), it remains an open
problem at the present time.

\acknowledgments

I am most grateful to G. Kotliar, M. Rozenberg, A. E. Ruckenstein,
and J. L. Smith for stimulating collaborations on this work,
and to E. Abrahams, J. Cardy, V. Dobrosavljevic, A. M. Finkelstein,
E. Fradkin, A. Georges, T. Giamarchi, K. Ingersent, A. J. Leggett,
I. Perakis, A. J. Schofield, A. Sengupta, C. Sire, C. M. Varma, YU Lu,
and X. Y. Zhang for useful discussions and/or communications.
I would like to acknowledge the hospitality of the 
Institute for Theoretical Physics, University of
California at Santa Barbara, and of the Aspen Center
for Physics. This research was supported in part by NSF Grant
No. PHY94-07194 at ITP, and by an A. P. Sloan Fellowship.

\newpage

\begin{figure}
\epsfxsize=6.5 in
\centerline{\epsffile{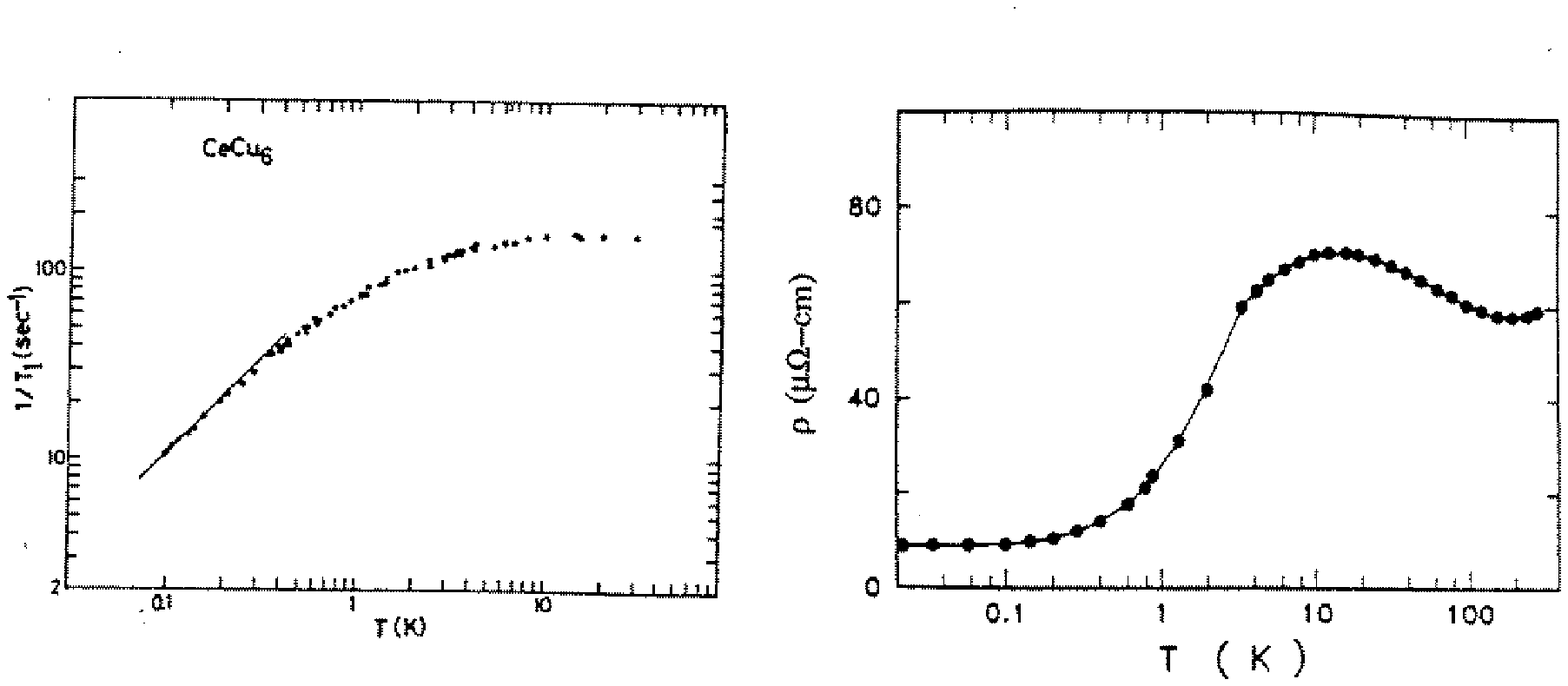}}
\vspace{0.25 in}
\caption{The NMR relaxation rate (${\rm 1/T_1}$) and electrical
resistivity ($\rho$) as a function of temperature in
the heavy fermion compound ${\rm CeCu_6}$.}
\label{fig:hf}
\end{figure}

\vskip 0.5in

\begin{figure}
\epsfxsize=6.5 in
\centerline{\epsffile{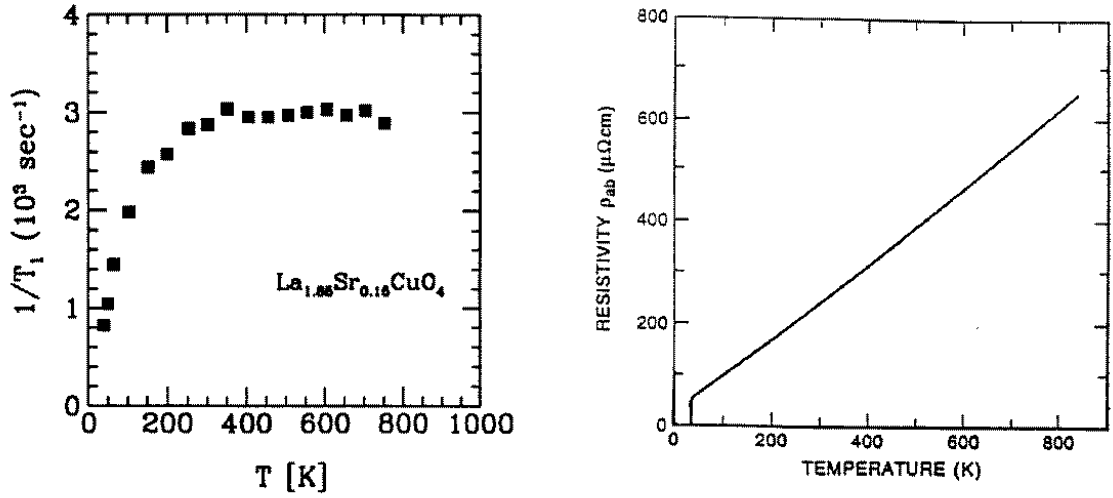}}
\vspace{0.25 in}
\caption{The NMR relaxation rate (${\rm 1/T_1}$) and electrical
resistivity in the ab-plane ($\rho_{ab}$) as a function
of temperature in the normal state of the high ${\rm T_c}$
compound  ${\rm La_{1.85}Sr_{0.15}CuO_4}$.}
\label{fig:cuprates}
\end{figure}

\newpage

\begin{figure}
\epsfxsize=2.5 in
\centerline{\epsffile{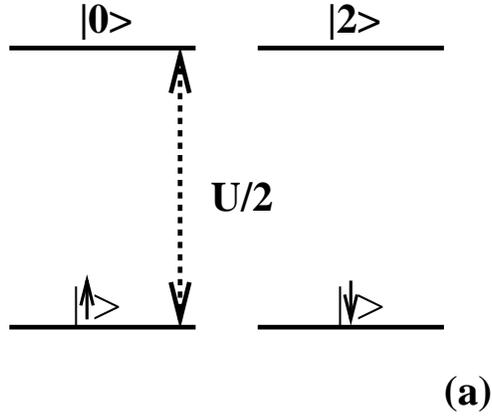}}
\vspace{1.0 in}
\epsfxsize=2.5 in
\centerline{\epsffile{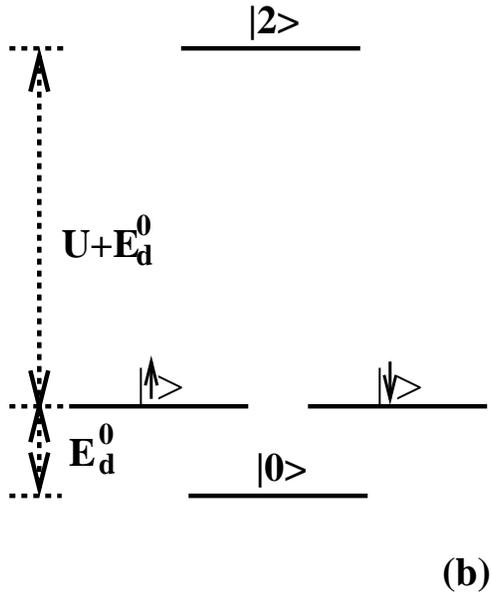}}
\vspace{0.5 in}
\caption{Impurity configurations and energy levels 
in (a) the symmetric Anderson model and (b) 
the asymmetric Anderson model.}
\label{fig:levels}
\end{figure}

\vskip 0.5in

\begin{figure}
\epsfxsize=3.5 in
\centerline{\epsffile{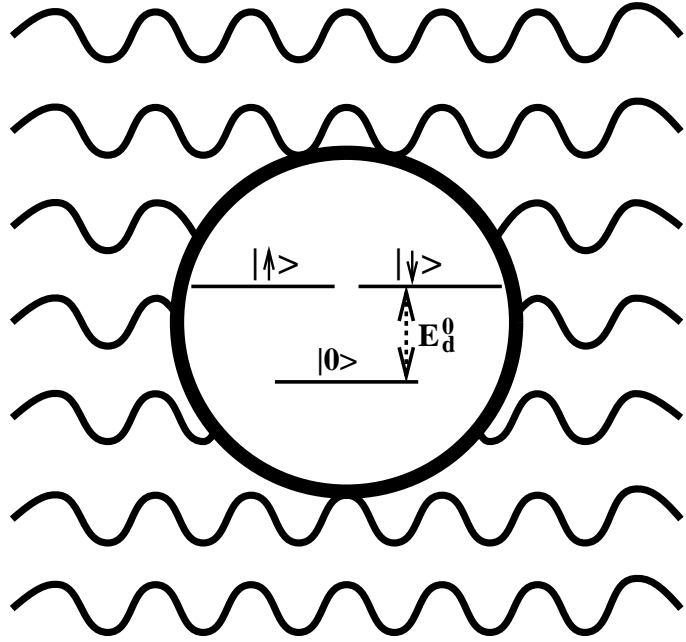}}
\vspace{0.5 in}
\caption{The generalized Anderson model as a three-level system. The
wavy lines represent the conduction electron bath.}
\label{fig:ThLS}
\end{figure}

\vskip 1.5in

\begin{figure}
\epsfxsize=5 in
\centerline{\epsffile{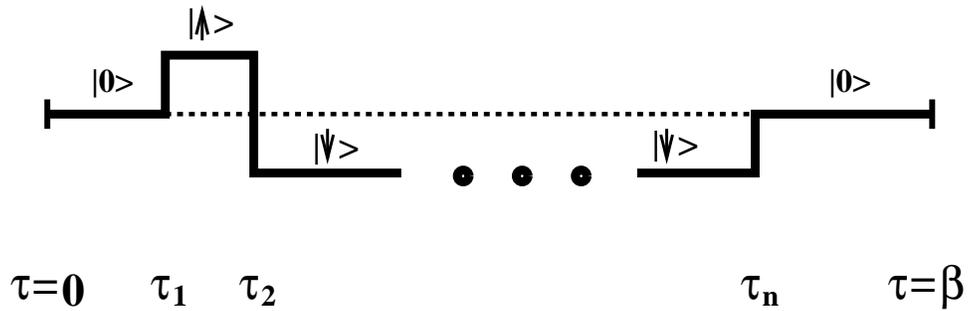}}
\vspace{0.5 in}
\caption{A typical hopping sequence in the atomic representation
along the imaginary time axis $\tau \in [0, \beta \equiv 1/T]$.
$\tau_i$, for $i=1,...,n$, labels the time at which the
impurity hops from one configuration to another.}
\label{fig:history}
\end{figure}

\vskip 0.5in
\newpage

\begin{figure}
\epsfxsize=3 in
\centerline{\epsffile{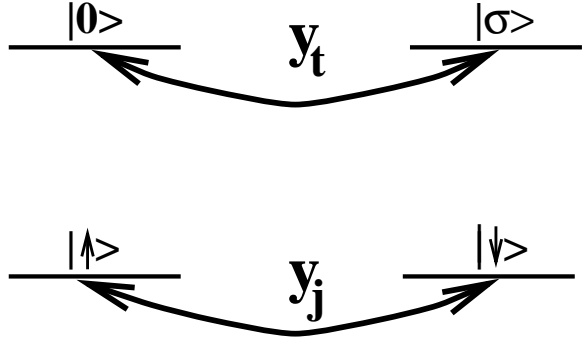}}
\vspace{0.25 in}
\caption{Schematic picture showing that the fugacities of the Coulomb
gas representation correspond to the dimensionless quantum transition 
amplitudes between the impurity configurations.  $y_t=t \xi_0$ is
the charge fugacity, and $y_j=J_{\perp}\xi_0$ the spin fugacity.
$\xi_0$ is the inverse energy cutoff.}
\label{fig:fugacity}
\end{figure}

\vskip 0.5in

\begin{figure}
\epsfxsize=5 in
\centerline{\epsffile{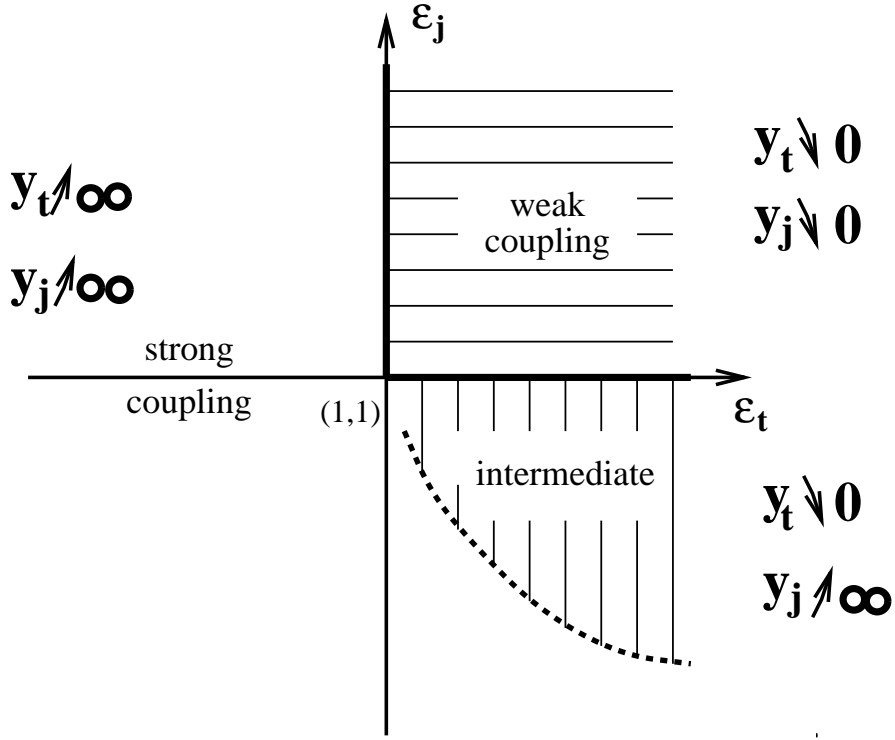}}
\vspace{0.25 in}
\caption{Phase diagram of the generalized Anderson model Eq. (\ref{hamiltonian.gam})
in the mixed valence regime. Here $\epsilon_t$ and $\epsilon_j$ label
the charge and spin stiffness constants defined in the text.
The vertical thick line, the horizontal thick line, and 
the dashed line are the boundaries between the different
mixed valence states. The dashed line is schematic.}
\label{fig:fixepoints}
\end{figure}

\vskip 0.5in

\begin{figure}
\epsfxsize=3 in
\centerline{\epsffile{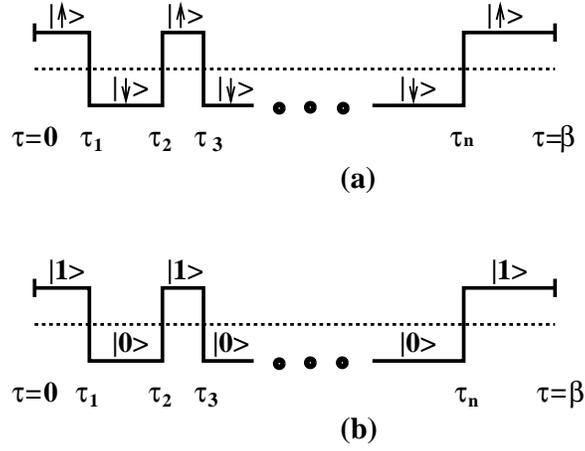}}
\vspace{0.25 in}
\caption{Hopping sequences in the atomic representation
for (a) the usual spin-Kondo problem and (b) the 
charge-Kondo problem, i.e., the resonant-level model.}
\label{fig:history2}
\end{figure}

\vskip 0.5 in

\begin{figure}
\epsfxsize=2.5 in
\centerline{\epsffile{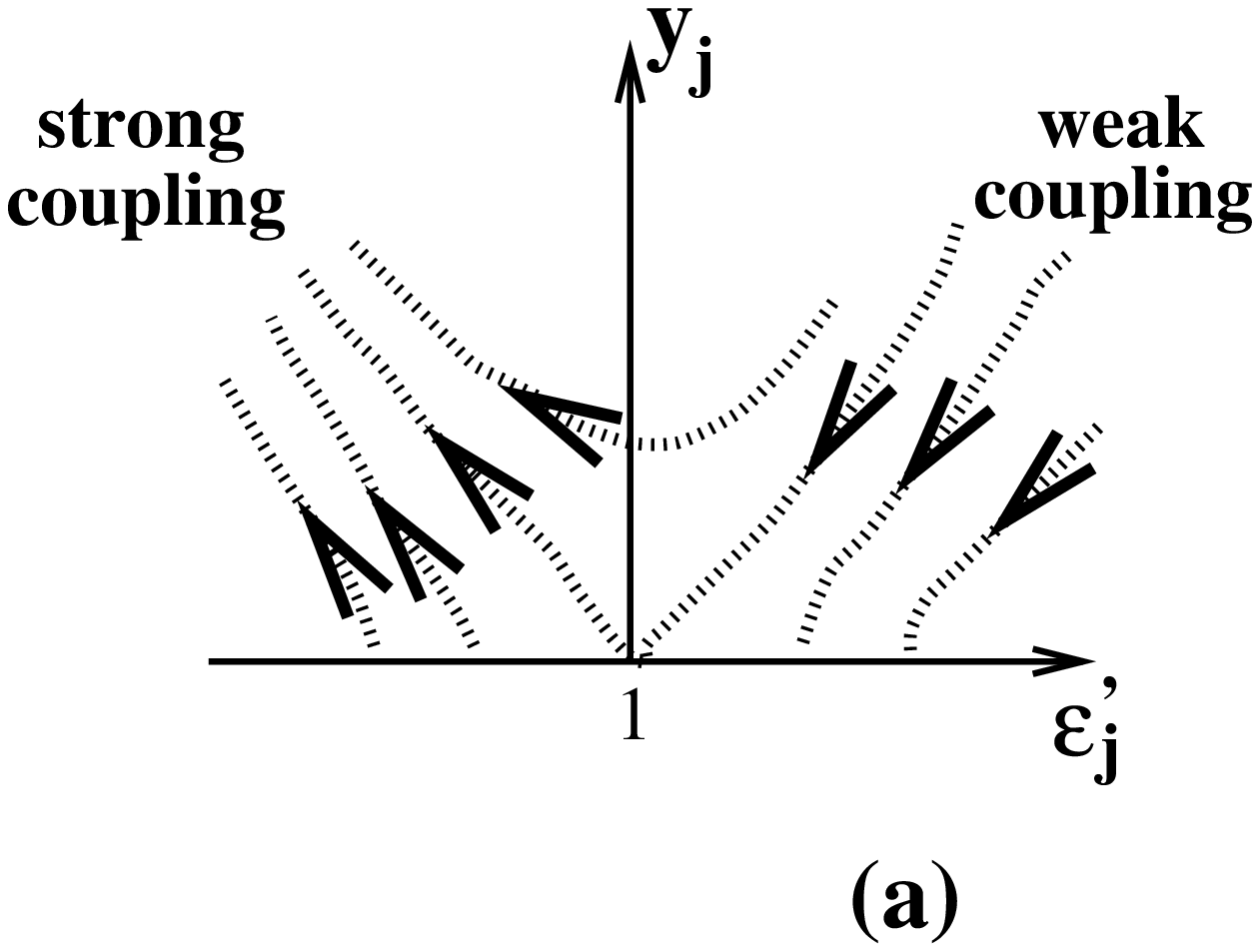}}
\vspace{0.25 in}
\epsfxsize=2.5 in
\centerline{\epsffile{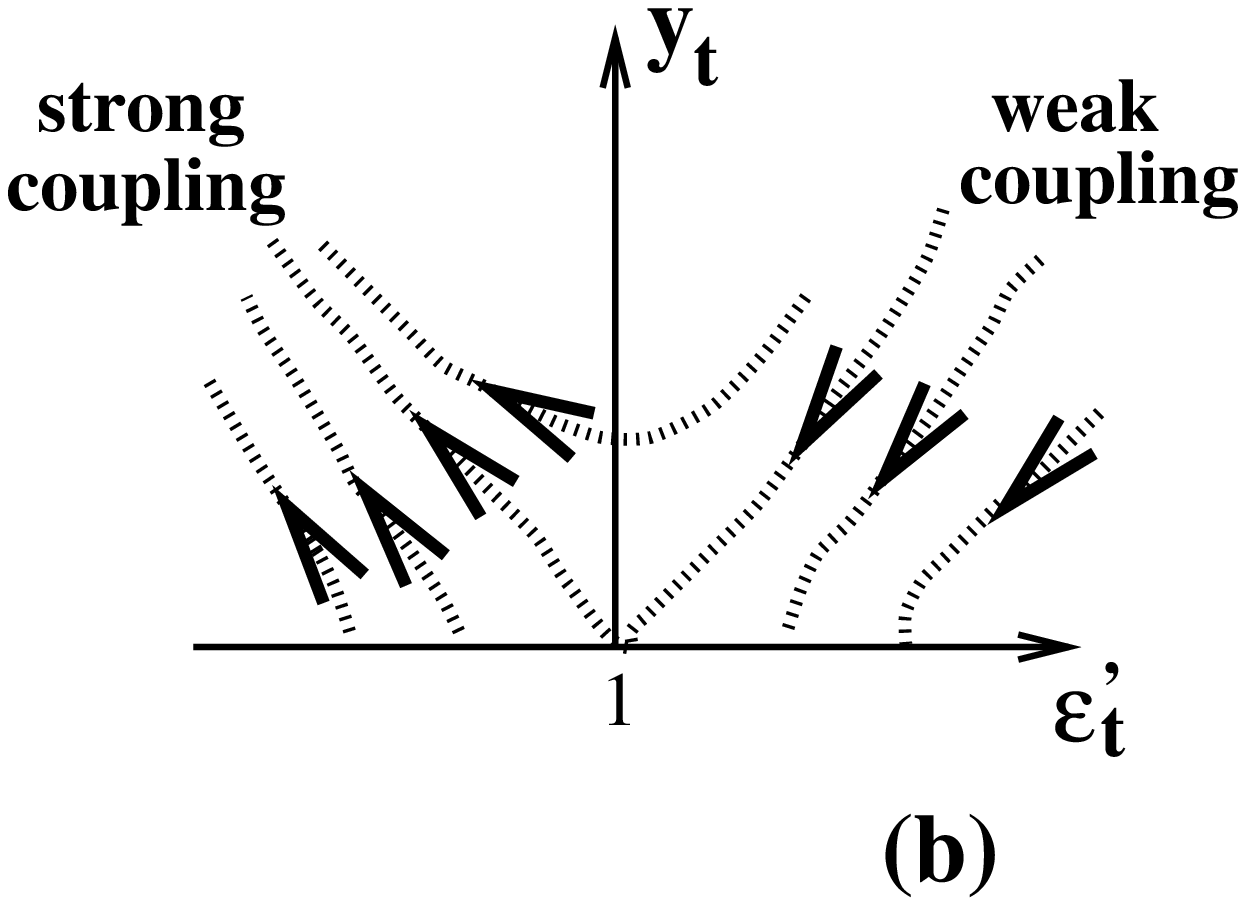}}
\vspace{0.25 in}
\caption{The RG flows for (a) the usual spin-Kondo problem and 
(b) the resonant-level model.}
\label{fig:flowsKondo}
\end{figure}

\vskip 0.5in
\newpage

\begin{figure}
\epsfxsize=2.0 in
\centerline{\epsffile{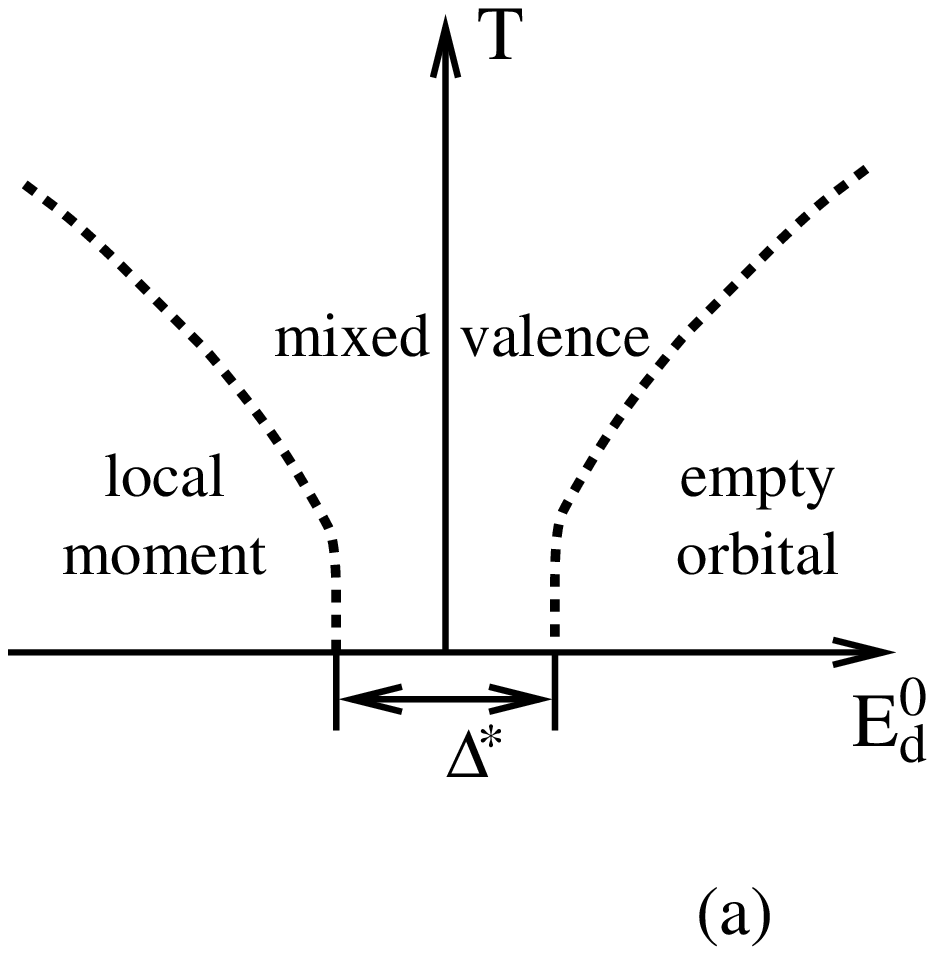}}
\vspace{0.25 in}
\epsfxsize=2.0 in
\centerline{\epsffile{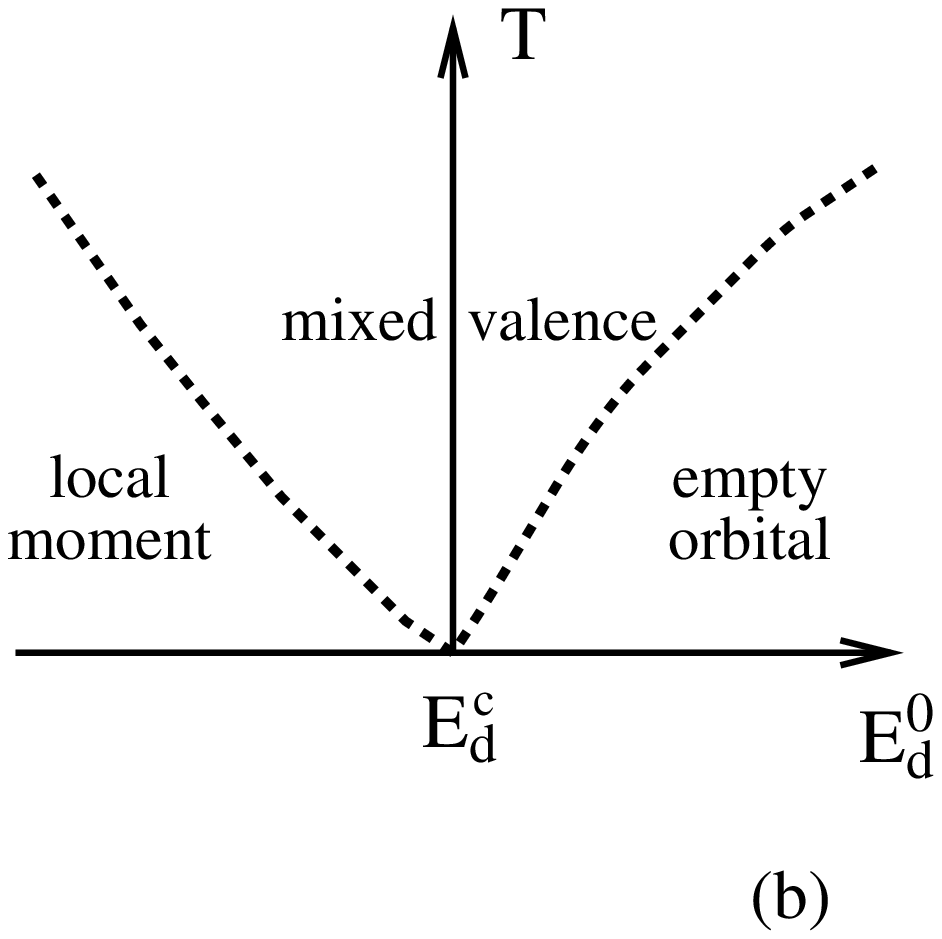}}
\vspace{0.25 in}
\caption{Crossover diagrams in terms of temperature ($T$) and
the impurity-level ($E_d^0$) (a) for the strong coupling phase
where $\Delta^*$ is the renormalized resonance width; and (b)
for the intermediate phase where $E_d^c$ labels the critical
impurity level.}
\label{fig:crossover-imp}
\end{figure}

\vskip 0.25in

\begin{figure}
\epsfxsize=4 in
\centerline{\epsffile{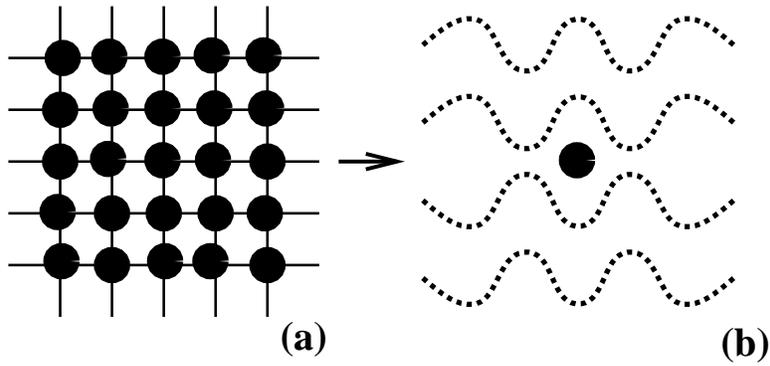}}
\vspace{0.25 in}
\caption{Schematic picture of the lattice model and its reduction to
an effective impurity model with self-consistent conduction electron
bath in the large $D$ limit.}
\label{fig:lattice}
\end{figure}

\newpage

\begin{figure}
\epsfxsize=3.5 in
\centerline{\epsffile{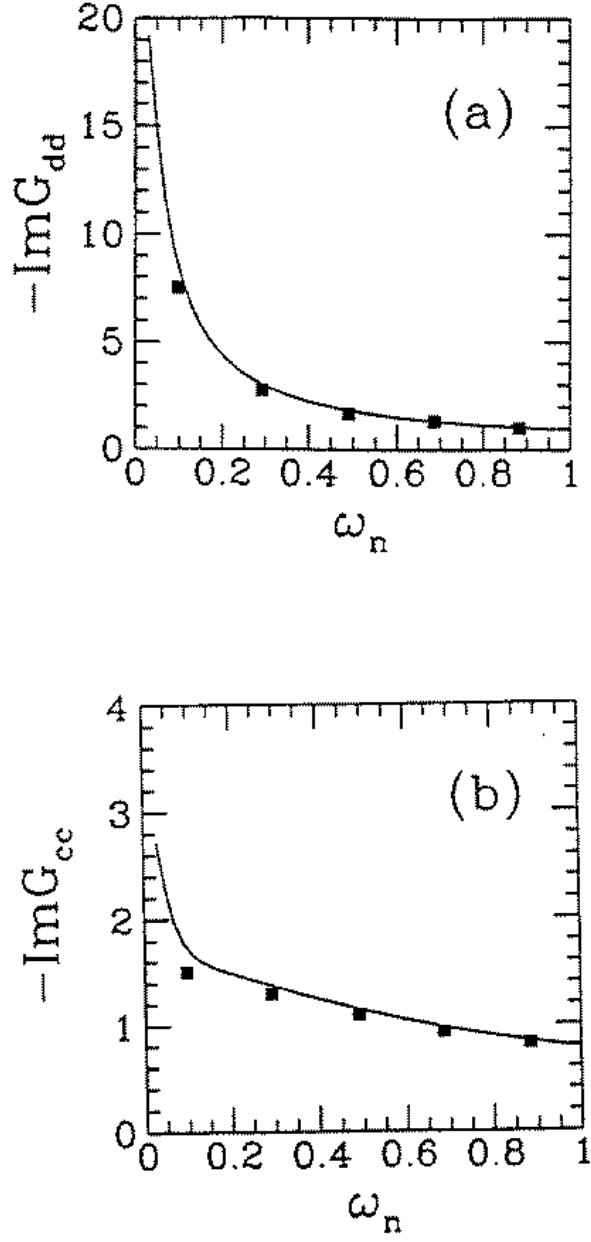}}
\vspace{0.5 in}
\caption{Numerical results of the $d-$ and $c-$ electron
Green's functions vs. the Matsubara frequency $\omega_n$
for a set of parameters for which the solution is 
a non-Fermi liquid metallic state. Solid lines come from the
self-consistent exact diagonalization method
and the solid squares are from quantum Monte Carlo
with $\beta=64$.}
\label{fig:numerics}
\end{figure}

\vskip 0.5in
\newpage

\begin{figure}
\epsfxsize=5.0 in
\centerline{\epsffile{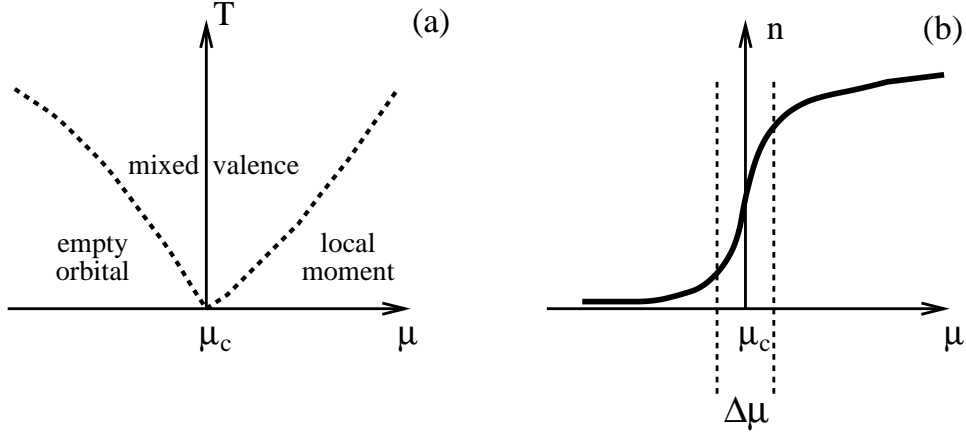}}
\vspace{0.25 in}
\caption{(a)Crossover in terms of temperature ($T$) and 
chemical potential ($\mu$) for the intermediate phase;
(b) Electron density ($n$) versus chemical potential
for the intermediate phase. $\Delta \mu \sim
( T )^{(2\epsilon_t^*-1)}$ where $\epsilon_t^*>1$ is
the renormalized charge stiffness constant.}
\label{fig:crossover-lattice}
\end{figure}

\vskip 0.5in

\begin{figure}
\epsfxsize=5.0 in
\centerline{\epsffile{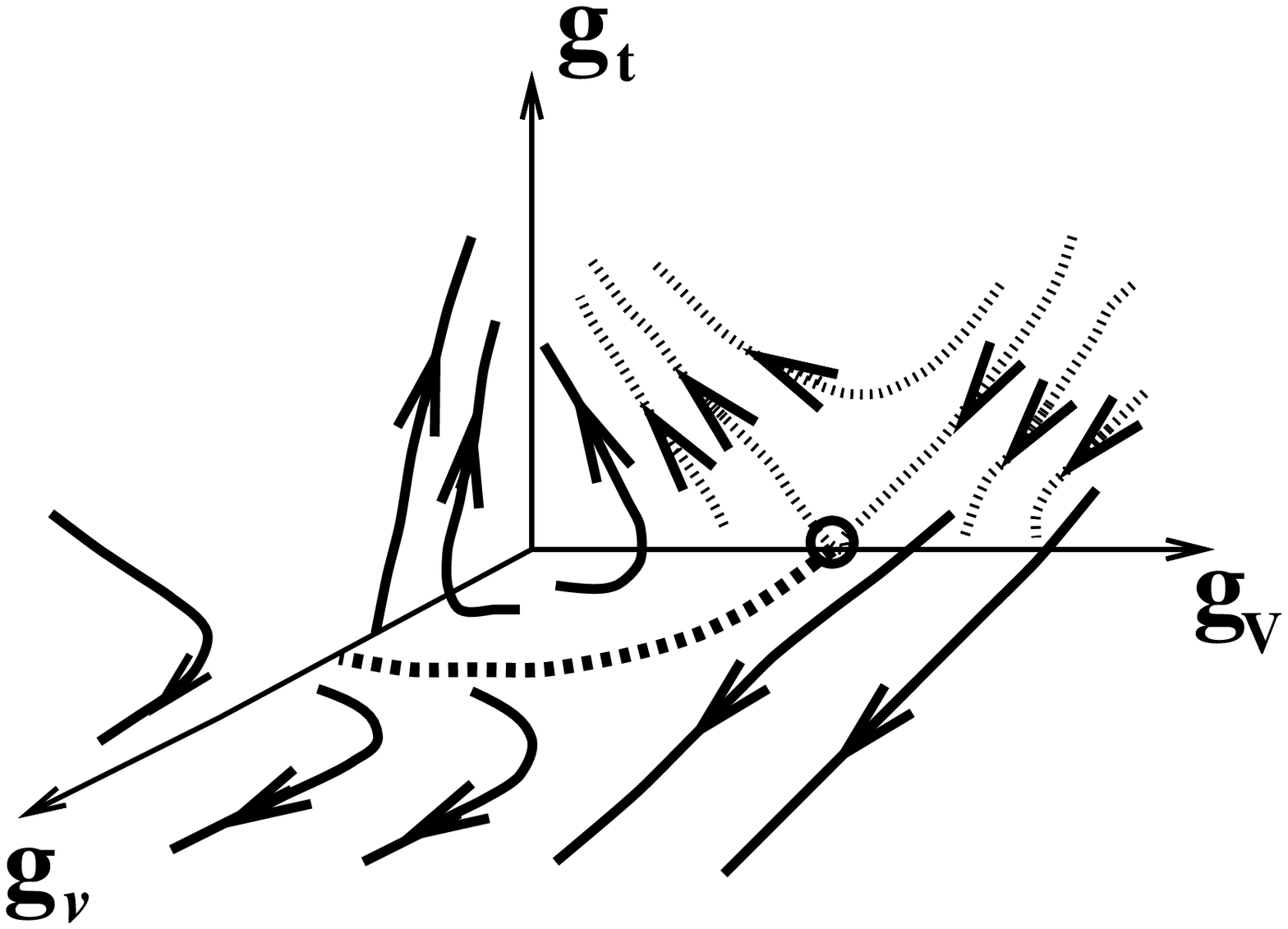}}
\vspace{0.25 in}
\caption{
The phase diagram of the Hamiltonian Eq. (\ref{hamiltonian.tv})
in the mixed valence regime. 
$g_t = t\rho_0$, $g_V = [1 - (2/\pi){tan^{-1}(\pi \rho_{0} V/2)}]$,
and $g_v = \rho_0 v_0$. The circle labels a Kosterlitz-Thouless
critical point. The dashed line is schematic. The phases are
described in the text.}
\label{fig:1/d}
\end{figure}


\begin{references}

\bibitem{Landau}L. D. Landau, Sov. Phys. JETP 3, 920 (1956);
5, 101 (1957); 8, 70 (1959).

\bibitem{Wheatley} J. Wheatley,  Rev. Mod. Phys. 47, 415 (1975).

\bibitem{LeggettHe3} A.J. Leggett, Rev. Mod. Phys. 47, 331 (1975).

\bibitem{McWhan} D. B. McWhan and T. M. Rice, Phys. Rev. Lett.
22, 887 (1969).

\bibitem{BrinkmanRice}W. F. Brinkman and T. M. Rice, 
Phys. Rev. B2, 4302 (1970).
 
\bibitem{mott} N. F. Mott, {\it Metal-Insulator Transitions}
(Taylor \& Francis, London, 1990).

\bibitem{Steglich} N. Grewe, F. Steglich, in {\it 
Handbook on the Physics and Chemistry of Rare Earths} Vol. 14,
eds. K. A. Gschneidner Jr. and L. Eyring (Elsevier,
Amsterdam, 1991), p. 343.

\bibitem{Leehf} P.A. Lee, T.M. Rice, J.W. Serene, L.J. Sham, and J.W. Wilkins,
Comm. Cond. Matt. Phys. 12, 99 (1986).

\bibitem{Houston} For latest developments, see
{\it Proceedings of 10th Anniversary HTS Workshop on Physics, 
Materials and Applications, Houston, TX, March 12-16, 1996},
editors B. Batlogg, W. K. Chu, and D. Gubser (World  Scientific,
Singapore, 1996).

\bibitem{Maple} M. B. Maple, C. L. Seaman, D. A. Gajewski,
Y. Dalichaouch, V. B. Barbetta, M. C. de Andrade,
H. A. Mook, H. G. Lukefahr, O. O. Bernal, D. E. MacLaughlin,
J. Low Temp. Phys. 95, 225 (1994).

\bibitem{Lonzarich} C. Pfleiderer, G. J. McMullan and G. G. Lonzarich,
Physica B206 \& 207, 847 (1995); F. M. Grosche,
C. Pfleiderer, G. J. McMullan, G. G. Lonzarich, and 
N. R. Bernhoeft,
{\it ibid.}, 20 (1995); S. R. Julian, N. D. Mathur,
F. M. Grosche, and G. G. Lonzarich, preprint (1996).

\bibitem{Lohneysen} H. v. L\"ohneysen, T. Pietrus, G. Portisch,
H. G. Schlager, A. Schr\"oder, M. Sieck, and T. Trappmann,
Phys. Rev. Lett. 72, 3262 (1994);
B. Bogenberger and H. v. L\"ohneysen,
{\it ibid.} 74, 1016 (1995).

\bibitem{Schulz} D. Jerome and H. J. Schulz, Adv. Phys. 31, 299 (1982);
C. Bourbonnais, in
{\it Strongly Interacting Fermions and High $T_c$
Superconductivity}, eds. B. Doucot and J. Zinn-Justin
(Elsevier, North Holland, 1995).

\bibitem{Voit} J. Voit, Rep. Prog. Phys. 58, 977 (1995).

\bibitem{Ralph} D. C. Ralph, A. W. W. Ludwig, J. von Delft, and
R. A. Buhrman, Phys. Rev. Lett. 72, 1064 (1994); {\it ibid.} 75,
770 (1995).

\bibitem{Altshuler} N. S. Wingreen, B. L. Altshuler, and Y. Meir,
Phys. Rev. Lett. 75, 769 (1995).

\bibitem{Solyom} V. J. Emery, in {\it Highly Conducting
One-dimensional Solids}, Eds. J. T. Devreese {\it et al.} (Plenum, New
York, 1979); J. Solyom, Adv. Phys. 28, 201 (1979).

\bibitem{Haldane1D} F.D.M. Haldane, J. Phys. C14, 2585 (1981).

\bibitem{Shankar} R. Shankar, Rev. Mod. Phys. 66, 129 (1994).

\bibitem{Castellani} C. Castellani, C. Di Castro, and W. Metzner,
Phys. Rev. Lett. 72, 316 (1994).

\bibitem{Engelbrecht} J. R. Engelbrecht and M. Randeria,
Phys. Rev. Lett. 65, 1032 (1990); Phys. Rev. B45, 12419 (1992).

\bibitem{Anderson} P. W. Anderson, Phys. Rev. Lett. 64, 1839 (1990);
{\it ibid.} 65, 2306 (1990).

\bibitem{Hewson} A. C. Hewson, {\it The Kondo Problem to
Heavy Fermions} (Cambridge University Press, 1993).

\bibitem{Blandin} P. Nozieres and A. Blandin, J. Phys.
(Paris) 41, 193 (1980).

\bibitem{MetznerVollhardt} W. Metzner and D. Vollhardt,
Phys. Rev. Lett. 62, 324 (1989).

\bibitem{largeD.reviews} For a comprehensive and
updated review, see A. Georges, G. Kotliar, W. Krauth,
and M. J. Rozenberg, Rev. Mod. Phys. 68, 13 (1996).
Earlier reviews on this subject include E. Muller-Hartmann 
Int. Jour. Mod. Phys. 3, 2169 (1989); D.Vollhardt,
Physica B169, 277 (1991); A. Georges, G. Kotliar, 
and Q. Si, Int. Journ. Mod. Phys. B6, 705 (1992);
Th. Pruschke, M. Jarrell, and J. K. Freericks,
Adv. Phys. 44, 187 (1995).

\bibitem{Cox} D. Cox, Phys. Rev. Lett. 59, 1240 (1987);
M. Jarrell, H. Pang, D. L. Cox, and K. H. Luk,
{\it ibid.} 77, 1612 (1996).

\bibitem{Kitaoka} K. Asayama, Y. Kitaoka, and Y. Kohori, 
J. Magn. Magn. Mat. 76\&77, 449 (1988).

\bibitem{Penney} T. Penney, F.P. Milliken,
S. von Molnar, F. Holtzberg, and Z. Fisk, Phys. Rev. B34, 5959 (1986).

\bibitem{form-factor} Presumably the dominant contribution to
$(1 /T_1)_{Cu}$ comes from the $f-$electron spin fluctuations;
there is no symmetry reason for the latter to be filtered out
due to special form factors of the hyperfine coupling.

\bibitem{YuvalAnderson} P. W. Anderson, G. Yuval and D.R. Hamman,
Phys. Rev. B1, 4464 (1970).

\bibitem{Wilson} K. G. Wilson, Rev. Mod. Phys. 47, 773 (1975).

\bibitem{AndreiWiegmann}N. Andrei, K. Furuya, and J. H. Lowenstein, 
Rev. Mod. Phys. 55, 331 (1983); A. M. Tsvelick and P. B. Wiegmann,
Adv. Phys. 32, 453 (1983).

\bibitem{Miranda} E. Miranda, V. Dobrosavljevic, and G. Kotliar,
this volume (1996).

\bibitem{Aronson} M. C. Aronson,
R. Osborn,  R. A. Robinson, J. W. Lynn,
R. Chau, C. L. Seaman, and M. B. Maple,
Phys. Rev. Lett. 75, 725 (1995); 
M. C. Aronson, M. B. Maple, P. de Sa, A. M. Tsvelik, and R. Osborn, 
preprint (1996).

\bibitem{Imai}T. Imai, C.P. Slichter, K. Yoshimuza, and K.
Kosuge, Phys. Rev. Lett. 70, 1002 (1993).

\bibitem{Batlogg}H. Takagi, B. Batlogg, H. L. Kao,
J. Kwo, R. J. Cava, J. J. Krajewski, and W. F. Peck, Jr.,
Phys. Rev. Lett. 69, 2975 (1992).

\bibitem{Hayden96} S. M. Hayden, G. Aeppli,
H. A. Mook, T. G. Perring, T. E. Mason,
S.-W. Cheong, and Z. Fisk, Phys. Rev. Lett. 76, 1344 (1996);
Q. Si, Y. Zha, and K. Levin, J. Appl. Phys. 76, 6935 (1994).

\bibitem{SKLL} Q. Si, J.H. Kim, J. P. Lu and K. Levin,
Phys. Rev. B42, 1033 (1990).

\bibitem{Bonn} D. A. Bonn, R. Liang, T. M. Riseman, D. J. Baar,
D. C. Morgan, K. Zhang, P. Dosanjh, T. L. Duty, A. MacFarlane,
G. M. Morris, J. H. Brewer, C. Kallin, and A. J. Berlinsky,
Phys. Rev. B47, 11314 (1993).

\bibitem{Ong} K. Krishana, J. M. Harris, and N. P. Ong,
Phys. Rev. Lett. 75, 3529 (1995).

\bibitem{Emery} V. Emery, Phys. Rev. Lett. 58, 2794 (1987).

\bibitem{VSA} C. M. Varma, S. Schmitt-Rink, and E. Abrahams,
Solid State Commun. 62, 681 (1987).

\bibitem{note:ehm} 
This model is not to be confused with the one-band extended
Hubbard model. The latter refers to the standard Hubbard model
generalized to include nearest-neighbor or further neighbor 
interactions.

\bibitem{SiKotliar1}
Q. Si and G. Kotliar, Phys. Rev. Lett. 70, 3143 (1993).

\bibitem{SiKotliar2}
Q. Si and G. Kotliar, Phys. Rev. B48, 13881 (1993).

\bibitem{KotliarSi1}
G. Kotliar and Q. Si, Phys. Scrip. T49, 165 (1993).

\bibitem{KotliarSi2}
G. Kotliar and Q. Si, Phys. Rev. B53, 12373 (1996).

\bibitem{Perakis}I. Perakis, C. M. Varma, and A. E. Ruckenstein,
Phys. Rev. Lett. 70, 3467(1993).

\bibitem{Giamarchi} T. Giamarchi, C. M. Varma, A. E. Ruckenstein, and
P. Nozieres, Phys. Rev. Lett. 70, 3967 (1993).

\bibitem{Yulu} G. M. Zhang and Lu Yu, Phys. Rev. Lett. 72, 2474;
G. M. Zhang, Z. B. Su, and Lu Yu, Phys. Rev. B53, 715 (1996).

\bibitem{Sire} C. Sire, C. M. Varma, A. E. Ruckenstein, and T. Giamarchi,
Phys. Rev. Lett. 72, 2478 (1994).

\bibitem{Costi} T. A. Costi, preprint (1996).

\bibitem{SRKR} Q. Si, M. Rozenberg, G. Kotliar, and A. E. Ruckenstein,
Phys. Rev. Lett. 72, 2761 (1994).

\bibitem{Smith} Q. Si and J. L. Smith, Phys. Rev. Lett. 77, in press
(available from cond-mat/9606087) (1996).

\bibitem{Coleman} P. Coleman, Phys. Rev. B29, 3035 (1984).

\bibitem{Read} N. Read and D. M. Newns, J. Phys. C16, 3273 (1983).

\bibitem{AuerbachLevin}
A. Auerbach and K. Levin, Phys. Rev. Lett. 57, 877 (1986).

\bibitem{MillisLee} A.J. Millis and P.A. Lee,
Phys. Rev. B35, 3394 (1987).

\bibitem{FalicovKimball} L. M. Falicov and J. C.
Kimball, Phys. Rev. Lett. 22, 997 (1969).

\bibitem{SchriefferWolff} J. R. Schrieffer and P. A. Wolff,
Phys. Rev. 149, 491 (1966).

\bibitem{AffleckLudwig}
I. Affleck and A. W.W. Ludwig, Nucl. Phys.
B360,  641 (1991).

\bibitem{VarmaYafet} C. M. Varma and Y. Yafet, Phys. Rev. B13, 
2950 (1976); C. M. Varma, Rev. Mod. Phys. 48, 219 (1976).

\bibitem{Haldane} 
F.D.M. Haldane, Phys. Rev. Lett. 40,416(1978);
J. Phys. C11, 5015 (1978).

\bibitem{Krishnamurthy}
H. R. Krishnamurthy,
K. G. Wilson and J. W. Wilkins, Phys. Rev. B21, 1044 (1980).

\bibitem{manganites} For recent theoretical discussions see, for 
example, N. Furukawa, J. Phys. Soc. Jpn. 63, 3214 (1994);
A. J. Millis, P. B. Littlewood, and B. I. Shraiman,
Phys. Rev. Lett. 74, 5144 (1995);
H. Roder, J. Zang, and A. R. Bishop, {\it ibid.} 76, 1356 (1996);
S. Sarkar, preprint (1996).

\bibitem{Haldane.srn} F. D. M. Haldane, Phys. Rev. B15, 2477 (1977).

\bibitem{Hakim} F. Guinea, V. Hakim, and A. Muramatsu, Phys. Rev.
B32, 4410 (1985).

\bibitem{AeppliFisk}
G. Aeppli and Z. Fisk, Comments Cond. Mat. Phys. 16,
155 (1992).

\bibitem{UedaRice} T. M. Rice and K. Ueda, Phys. Rev. Lett. 55,
995 (1985).

\bibitem{Brandow} B. H. Brandow, Phys. Rev. B37, 250 (1988).

\bibitem{Grilli}M. Grilli, R. Raimondi, C. Castellani, C. Di Castro,
and G. Kotliar, Phys. Rev. Lett. 66, 1236 (1990).

\bibitem{Cardy} J. L. Cardy, J. Phys. A14, 1407 (1981).
See also S. Chakravarty and J. Hirsch, Phys. Rev. B 25,
3273 (1982).

\bibitem{Leggett}A. J. Leggett, 
S. Chakravarty, A. T. Dorsey, M.P.A. Fisher, 
and W. Zwerger, Rev. Mod. Phys. 59, 1 (1987);
{\it ibid.} 67, 215 (Erratum) (1995).

\bibitem{Wiegmann}
P.B. Wiegmann and A.M. Finkelstein, Sov. Phys. JETP,
48, 102 (1978).

\bibitem{Schlottman}
 P. Schlottmann, Phys. Rev. B25, 4815 (1982).

\bibitem{KaneFisher}C. L. Kane and M. P. A. Fisher,
Phys. Rev. B46, 15233 (1992).

\bibitem{Sachdev}For a review, see S. Sachdev,
{\it Proceedings of the 19th IUPAP International Conference
on Statistical Physics, Xiamen, China, 1995},
ed. B.-L. Hao (World Schientific, to be published).

\bibitem{Furusaki} A. Furusaki and N. Nagaosa, Phys. Rev. B47,
4631 (1993).

\bibitem{Nozieres} P. Nozieres, J. Low Temp. Phys. 17, 31 (1974).

\bibitem{Bethe} While the Kondo problem and the usual Anderson model
with infinite bandwidth are exactly soluble using the Bethe 
ansatz method\cite{AndreiWiegmann} and conformal field
theory\cite{AffleckLudwig}, the Anderson model with an 
additional density-density interaction has not been solved this way.

\bibitem{Toulouse} G. Toulouse, C. R. Acad. Sci. 268, 1200 (1969).

\bibitem{Heid} R. Heidenreich, R. Seiler, and 
D. A. Uhlenbrock, J. Stat. Phys. 22, 27 (1980) and references
therein.

\bibitem{Neuberg} H. Neuberger, Tel Aviv University Thesis (1975).

\bibitem{Blume} M. Blume, V. J. Emery, and A. Luther, 
Phys. Rev. Lett. 26, 1547 (1971); V. J. Emery and A. Luther,
Phys. Rev. B9, 215 (1974).

\bibitem{skg} Q. Si, G. Kotliar, and A. Georges, Phys. Rev. B46,
1261 (1992).

\bibitem{ZhangRice} F. C. Zhang and T. M. Rice,
Phys. Rev. B37, 3759 (1988).

\bibitem{EmeryReiter} V. J. Emery and G. Reiter,
Phys. Rev. B41, 7247 (1990).

\bibitem{Jarrell} M. Jarrell, Phys. Rev. Lett. 69, 168 (1992).

\bibitem{Rozenberg} M. Rozenberg, X. Y. Zhang, and G. Kotliar,
Phys. Rev. Lett. 69, 1236 (1992).

\bibitem{GeorgesKrauth} A. Georges and W. Krauth,
Phys. Rev. Lett. 69, 1240 (1992).

\bibitem{Caffarel} M. Caffarel and W. Krauth, 
Phys. Rev. Lett. 72, 1545 (1994).

\bibitem{Jones} B. A. Jones, C. M. Varma, and J. W. Wilkins,
Phys. Rev. Lett. 61, 125 (1988).

\bibitem{Schiller} A. Schiller and K. Ingersent, Phys. Rev. Lett.
75, 113 (1995).
        
\bibitem{Vlad} V. Dobrosavljevic and G. Kotliar,
Phys. Rev. B50, 1430 (1994).

\bibitem{Kajueter} Kajueter and Kotliar have independently
constructed related mean field equations in the context of
a spinless one-band fermion model with semi-circular
density of states. G. Kotliar (private communications); 
H. Kajueter, Rutgers Ph. D. thesis (1996).

\bibitem{Fradkin} D. Withoff and E. Fradkin, Phys. Rev. Lett.
64, 1835 (1990); C. R. Cassanello and E. Fradkin,
SISSA cond-mat/9512064 (1995).

\bibitem{Ingersent} K. Ingersent, SISSA cond-mat/9605025 (1996).

\bibitem{Si} Q. Si, unpublished (1996).

\bibitem{Bhatt}J. Bhattacharjee, S. Chakravarty, J. L. Richardson,
and D. J. Scalapino, Phys. Rev. B24, 3862 (1981).

\bibitem{Imbrie} J. Z. Imbrie and C. M. Newman, Commun. Math. Phys.
118, 303 (1988).

\bibitem{Smith2} J. L. Smith and Q. Si, unpublished (1996).

\bibitem{Allen} J.-S. Kang, J. W. Allen, M. B. Maple, M. S. Torikachvili,
W. P. Ellis, B. B. Pate, Z.-X. Shen, J. J. Yeh, and I. Linday,
Phys. Rev. B39, 13529 (1989).

\bibitem{Maple.flt} M. B. Maple, D. A. Gajewski,
C. L. Seaman, and J. W. Allen,P
Physica B199 \& 200, 423 (1994).

\bibitem{Kastner} See, for example, 
M. Kastner, Physics Today 46, 24 (1993).

\bibitem{Spindiff} Q. Si, preprint cond-mat/9507050; 
Q. Si, in Ref. \onlinecite{Houston} (1996).

\end{references}
\end{document}